\documentclass[a4paper,11pt]{article}
\usepackage{jinstpub} 
\usepackage{lineno}
\usepackage{subcaption}
\usepackage{graphicx}
\usepackage[symbol]{footmisc}

\title{Burn-in Test and Thermal Performance Evaluation of Silicon Photomultipliers for the JUNO-TAO Experiment}

\author{X. Chen$^{\mathrm{a}}$, G.F. Cao$^{\mathrm{b},\mathrm{c}}$, M.H. Qu$^{\mathrm{d}}$, H.W. Wang$^{\mathrm{b},}$\footnote[1]{Corresponding author.}, N. Anfimov$^{\mathrm{e}}$, A. Rybnikov$^{\mathrm{e}}$, J.Y. Xu$^{\mathrm{b},\mathrm{c}}$, A.Q. Su$^{\mathrm{d}}$, Z.L. Chen$^{\mathrm{a}}$, J. Cao$^{\mathrm{b}}$, Y.C. Li$^{\mathrm{b}}$, M. Qi$^{\mathrm{a}}$}
\affiliation{$^{a}$Nanjing University,\\
No.22 Hankou Road, Gulou District,  Nanjing, Jiangsu 210093, China}
\affiliation{$^{b}$Institute of High Energy Physics, Chinese Academy of Sciences,\\
No.19B Yuquan Road, Shijingshan District, Beijing 100049, China}
\affiliation{$^{c}$University of Chinese Academy of Sciences, \\
No.1 Yanqihu East Rd, Huairou District, Beijing 101408, China}
\affiliation{$^{d}$Zhengzhou University,\\
No.100 Science Avenue, Zhengzhou, Henan 450001, China}
\affiliation{$^{e}$Joint Institute for Nuclear Research,\\
6 Joliot-curie str., Dubna, Moscow Region 141980, Russia}
\emailAdd{wanghanwen@ihep.ac.cn}

\abstract{
This study evaluates more than 4,000 tiles made of Hamamatsu visual-sensitive silicon photomultipier (SiPM), each with dimensions of 5 $\times$ 5 cm$^2$, intended for the central detector of the Taishan Anti-neutrino Observatory (TAO), a satellite experiment of the Jiangmen Underground Neutrino Observatory (JUNO) aimed at measuring the reactor anti-neutrino energy spectrum with unprecedented energy resolution. All SiPM tiles underwent a room temperature burn-in test in the dark for two weeks, while cryogenic testing analyzed the thermal dependence of parameters for some sampled SiPMs. Results from these comprehensive tests provide crucial insights into the long-term performance and stability of the 10 square meters of SiPMs operating at -50°C to detect scintillation photons in the TAO detector. Despite some anomalies awaiting further inspection, all SiPMs successfully passed the burn-in test over two weeks at room temperature, which is equivalent to 6.7 years at -50°C. Results are also used to guide optimal SiPM selection, configuration, and operation, ensuring reliability and sustainability in reactor neutrino measurements. This work also provides insights for a rapid and robust quality assessment in future experiments that employ large-scale SiPMs as detection systems.}

\keywords{Neutrino, Photon detectors, SiPM, Cryogenic detectors}

\arxivnumber{2406.12912} 

\begin{document}
\maketitle
\flushbottom

\section{Introduction}
\label{sec:intro}
The Taishan Antineutrino Observatory (TAO, also known as JUNO-TAO) is a satellite experiment of the Jiangmen Underground Neutrino Observatory (JUNO) \cite{juno2016}. It is a next-generation reactor neutrino experiment aimed at precise measurements of the neutrino energy spectrum and the exploration of potential new physics \cite{junocollaboration2020tao}. The TAO experiment is designed to position its detector, comprising 2.8 tons of Gadolinium-doped Liquid Scintillator (GdLS) and 4,024 Silicon Photomultiplier (SiPM) tiles covering a spherical area of nearly 10 $\mathrm{m}^2$, in close proximity to one of the Taishan nuclear reactors. SiPMs are crucial components of the detector system, responsible for detecting faint scintillation light from neutrino interactions with a high photon detection efficiency of around 50\%.

A number of dark matter and neutrino experiments, notably including DarkSide-20k \cite{aalseth_darkside-20k_2018} and nEXO \cite{nexocollaboration2018nexopreconceptualdesignreport}, utilize liquid-phase noble gases and employ large area of SiPMs serving as photo-detectors under cryogenic conditions. Significant progress has been achieved in detector development and SiPM testing, as documented in various studies \cite{lAr-nima, Wang_2021, Baudis_2018, GALLINA2019371, nexo-2022-epjc, Balmforth_2023}.
To mitigate the dark noise in the SiPMs, the TAO detector is designed to operate at a temperature of -50°C (223K), adopting a new low-temperature liquid scintillator technology \cite{XIE2021165459}. Given the critical importance of long-term sustainability, it is crucial to conduct studies on the thermal behavior of SiPMs, with a particular emphasis on burn-in tests. These tests are of great importance for detecting early failures and ensuring that the SiPMs' performance remains consistent over time, thus reinforcing the detector's reliability.

In our study, we subjected the SiPMs to a rigorous burn-in test at room temperature to comprehensively evaluate their stability and reliability. We recorded dark current and temperature over an extended period of approximately two weeks to identify potential failures and signs of degradation. Additionally, we systematically examined various SiPM parameters, such as breakdown voltage, dark count rate (DCR), and crosstalk, among others across different temperatures. These parameters were studied under cryogenic conditions within a specialized low-temperature chamber.

The dark current, denoted as $I_{\mathrm{dark}}$, is a critical parameter in assessing the performance and reliability of SiPMs during the burn-in test. Our investigations have established that $I_{\mathrm{dark}}$ can be comprehensively represented as a function of temperature ($T$). This relationship stems from the fundamental insight that key factors influencing $I_{\mathrm{dark}}$, such as the dark count rate (DCR), crosstalk ($\lambda$), and pixel gain ($G_{\mathrm{pixel}}$), depend on temperature either directly or through their reliance on the temperature-sensitive breakdown voltage ($V_{\mathrm{bd}}(T)$), or both.
To accurately describe the dark current at room temperature, accounting for a small but non-negligible range of temperature variations, we use the following equation:
\begin{equation}
\label{eq:idark}
I_\mathrm{dark} = G_\mathrm{pixel} \cdot \frac{1}{1-\lambda} \cdot \mathrm{DCR} \cdot e,
\end{equation}
where $G_\mathrm{pixel}$ represents the gain of an individual SiPM pixel, $\lambda$ denotes the crosstalk parameter, $e$ is the elementary charge ($1.602 \times 10^{-19}$ C), $\mathrm{DCR}$ is the dark count rate, which can be further described by the temperature $T$ and overvoltage $V_\mathrm{ov}$:
\begin{equation}
\label{eq:dcr-ov}
\mathrm{DCR} = \mathcal{S}(T) \cdot (V_\mathrm{ov} - V_\mathrm{act}),
\end{equation}
\begin{equation}
V_\mathrm{ov} = V_\mathrm{bias} - V_\mathrm{bd}(T),
\end{equation}
where $\mathcal{S}(T)$ is the slope of the $\mathrm{DCR}$ versus overvoltage and it only depends on the temperature $T$  if we postulate a linear relationship between $\mathrm{DCR}$ and the overvoltage at a specified temperature, $V_\mathrm{act}$ represents the overvoltage threshold above which $\mathrm{DCR}$ begins to activate, $V_\mathrm{bias}$ is the bias voltage applied to the SiPM, and $V_\mathrm{bd}$ denotes the breakdown voltage. Furthermore, the breakdown voltage is subject to variations with temperature:

\begin{equation}
V_\mathrm{bd}(T) = V_\mathrm{ref} + (T - T_\mathrm{ref}) \cdot \frac{dV_\mathrm{bd}}{dT},
\end{equation}
in which $V_\mathrm{ref}$ and $T_\mathrm{ref}$ are reference values for the breakdown voltage and temperature, respectively. $V_\mathrm{ref}$ is determined with reference temperature $T_\mathrm{ref} = 24^{\circ}$C. We have also verified that there is a linear relation between $V_\mathrm{bd}$ and $T$ (see Fig.\,\ref{fig:vbd_temp} later on in the paper).

The pixel gain increases linearly with overvoltage as far as it remains inside the operating overvoltage range that we measured:
\begin{equation}
G_\mathrm{pixel} = C_\mathrm{cell} \cdot V_\mathrm{ov}
\end{equation}
where $C_\mathrm{cell}$ represents the capacitance of the SiPM cell. It is important to note that if $G_\mathrm{pixel}$ is expressed in number of electrons, then $C_\mathrm{cell}$ is considered in terms of capacitance; otherwise, it is a dimensionless coefficient.

The crosstalk parameter $\lambda$ can be also approximated to have a linear relation towards overvoltage $V_\mathrm{ov}$:
\begin{equation}
\begin{aligned}
    \lambda = &\; \kappa \cdot (V_\mathrm{ov} -\mathcal{V}_\mathrm{s})
\end{aligned}
\end{equation}
In this equation, $\kappa$ represents the slope, indicating the rate of change in $\lambda$ with overvoltage, and $\mathcal{V}_\mathrm{s}$ is the threshold overvoltage at which crosstalk begins to manifest.

It is important to note that the afterpulse effect is disregarded in Equation \ref{eq:idark}, as it only becomes significant at higher overvoltages (above 3.5 V) for the devices tested. The additional charge from afterpulsing is minimal; for instance, even at overvoltages of 6-7 V, it constitutes less than 2.5\% of the charge of a single photon-electron for most of the tested devices. All of the conclusions above are based on testing results at -50$^\circ$C. Our study shows that at higher temperatures, the afterpulse contribution decreases, making its impact even more negligible.

Finally, the dark current model is formulated as:
\begin{equation}
\label{eq:est-current}
I_\mathrm{dark}(T) = G_\mathrm{pixel}(V_\mathrm{ov}) \cdot \frac{1}{1-\lambda(V_\mathrm{ov})} \cdot \mathcal{S}(T) \cdot V_\mathrm{ov} \cdot e
\end{equation}

\section{Experimental Setup}
\label{sec:exp}
To facilitate the mass testing of the SiPMs for TAO, we have developed and implemented a series of experimental setups. This article focuses on two key tests: the thermal dependence test, examining how SiPM performance varies with temperature, and the burn-in test, which evaluates SiPM performance at room temperature.
\subsection{SiPMs for TAO}
\begin{figure}
    \centering
    \includegraphics[width=0.8\linewidth]{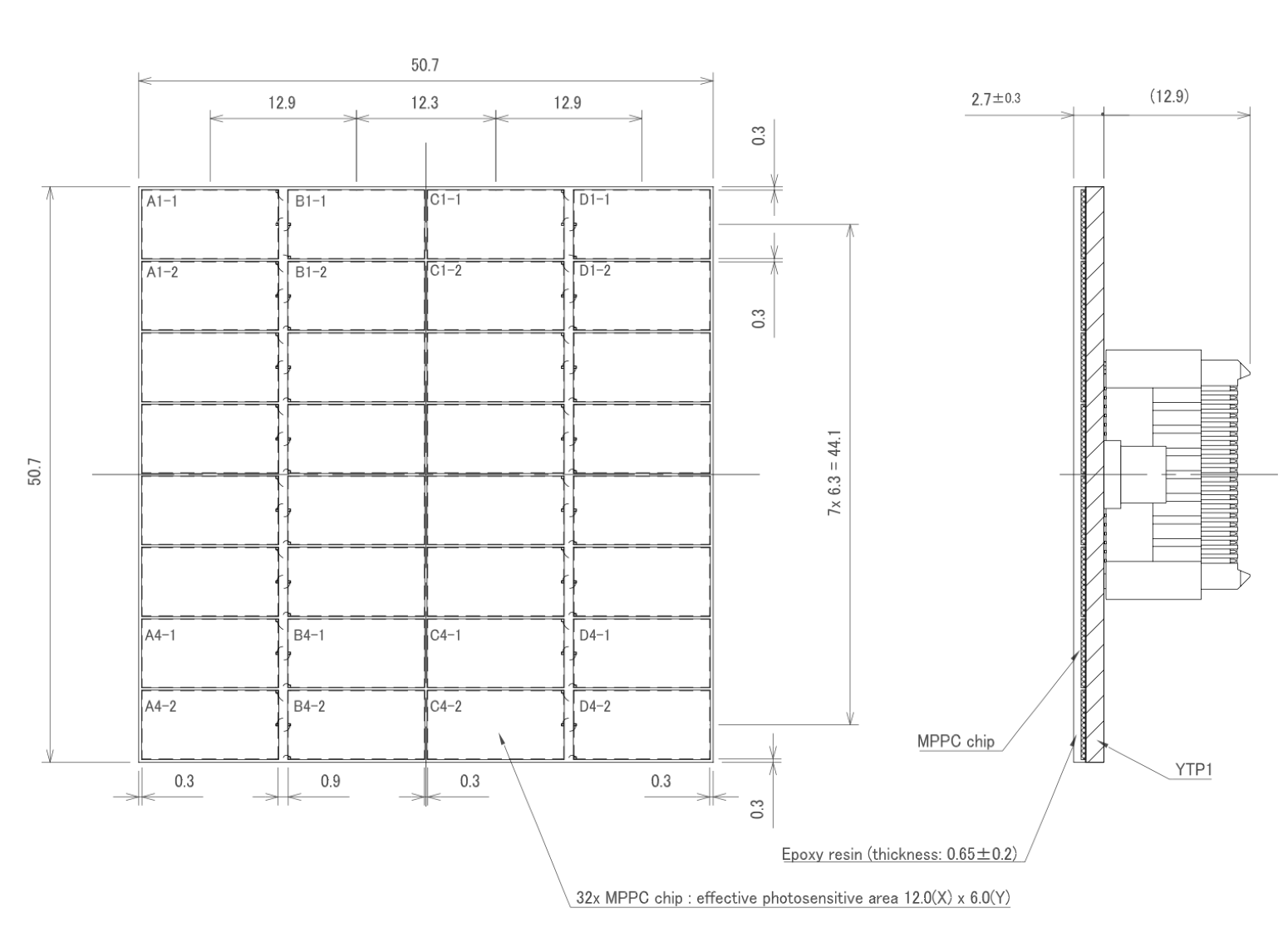}
    \caption{Dimensional outline of the front and side view of the S16088 SiPM tile}
    \label{fig:hpk-sipm-tile}
\end{figure}

We employ the Hamamatsu MPPC S16088 with structure details discussed in Table \ref{tab:hpk-sipm} and the design shown in Figure \ref{fig:hpk-sipm-tile} \cite{s16088}, which integrates thirty-two 6 $\times$ 12 mm$^2$ chips into a single tile. Each pair of adjacent chips, together covering an area of 12 $\times$ 12 mm$^2$, constitutes a single channel. The design features a pixel pitch of 75 $\mu$m, with each chip hosting 12,782 pixels, achieving a photon-sensitive area coverage of 89.6\%. Protecting the silicon sensor surface is a 0.65 $\pm$ 0.20 mm thick layer of epoxy resin with a refraction index of 1.54 for the blue light.
\begin{table}[htp]
    \centering
    \caption{Structure details of the HPK S16088 SiPM}
    \renewcommand{\arraystretch}{1.2}
    
    \begin{tabular}{c|c|c}
    \cline{1-3}
     Parameters & Value & Unit \\
    \cline{1-3} Number of channels & 16 $(4 \times 4)$ & - \\
     Effective photosensitive area & $12 \times 12$& $\mathrm{mm}^2 / \mathrm{ch}$. \\
     Coverage of photosensitive area & 89.6 & $\%$ \\
     Pixel pitch & 75 & $\mu \mathrm{m}$ \\
     Number of pixels / channel & 25,564& - \\
     Window & Epoxy resin & - \\
     Window refractive index & 1.54 & - \\
    \cline{1-3}
    \end{tabular}
   
    \label{tab:hpk-sipm}
\end{table}

Given the compact design of the SiPM tile, each measuring 50.7 mm $\times$ 50.7 mm and comprising 16 channels for efficient arrangement, the central detector of the TAO project employs a total of 4024 such SiPM tiles.

\subsection{The Characterization Test in a Cryogenic Chamber}
Our experimental setup includes a customized stand for recording SiPM waveforms in dark, low-temperature conditions \cite{Rybnikov_2024}.  
The testing platform resides within a cryogenic chamber, while the LED light source, power supply, and data acquisition (DAQ) system are located outside and connected via fibers and cables. The LED operates in pulse mode, generating weak illumination for the SiPMs. Additionally, we record waveforms under dark conditions to analyze the dark count rate of the SiPMs. This setup allows us to characterize the performance of SiPMs in a controlled, cryogenic environment, enabling precise measurements of their signal response and noise characteristics at low and varying temperatures. During each data acquisition session, referred to as a ‘run’, sixteen SiPM tiles underwent simultaneous evaluation at a certain temperature, specifically after fully cooling down, ensuring that the system had reached thermal equilibrium.

\subsection{The Burn-in Test}
\begin{figure}[htb]
    \centering
    \includegraphics[width=0.75\linewidth]{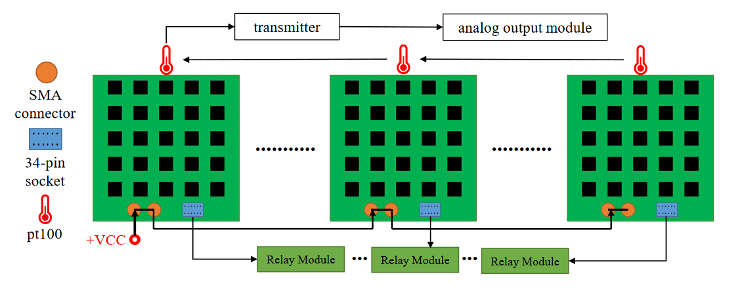}
    \caption{Experimental setup for the burn-in test featuring PCBs within a temperature-monitored dark chamber}
    \label{fig:burn-in-hardware}
\end{figure}
The configuration of the burn-in test is depicted in Figure \ref{fig:burn-in-hardware}, showcasing 16 large printed circuit boards (PCBs) positioned inside a dark chamber that is regulated at room temperature. This dedicated setup was designed and developed by another electronics team \cite{burn-in-article}. Each PCB is designed to accommodate 25 SiPM tiles, interconnected through Samtec connectors to facilitate easy and flexible replacement. The setup is equipped with PT100 sensors \cite{pt100} for continuous ambient temperature monitoring. Each tile on these boards is subjected to around 54 V voltage for a duration of two weeks. We precisely measure the current of each tile using a Keithley 6487 picoammeter, coupled with relay modules for effective management of the testing channels.

\section{Results and Discussions}
\label{sec:result}
\subsection{Characterizations at Low Temperatures}
\paragraph{Breakdown Voltage}
\begin{figure}[htb]
    \centering
    \includegraphics[width=0.5\linewidth]{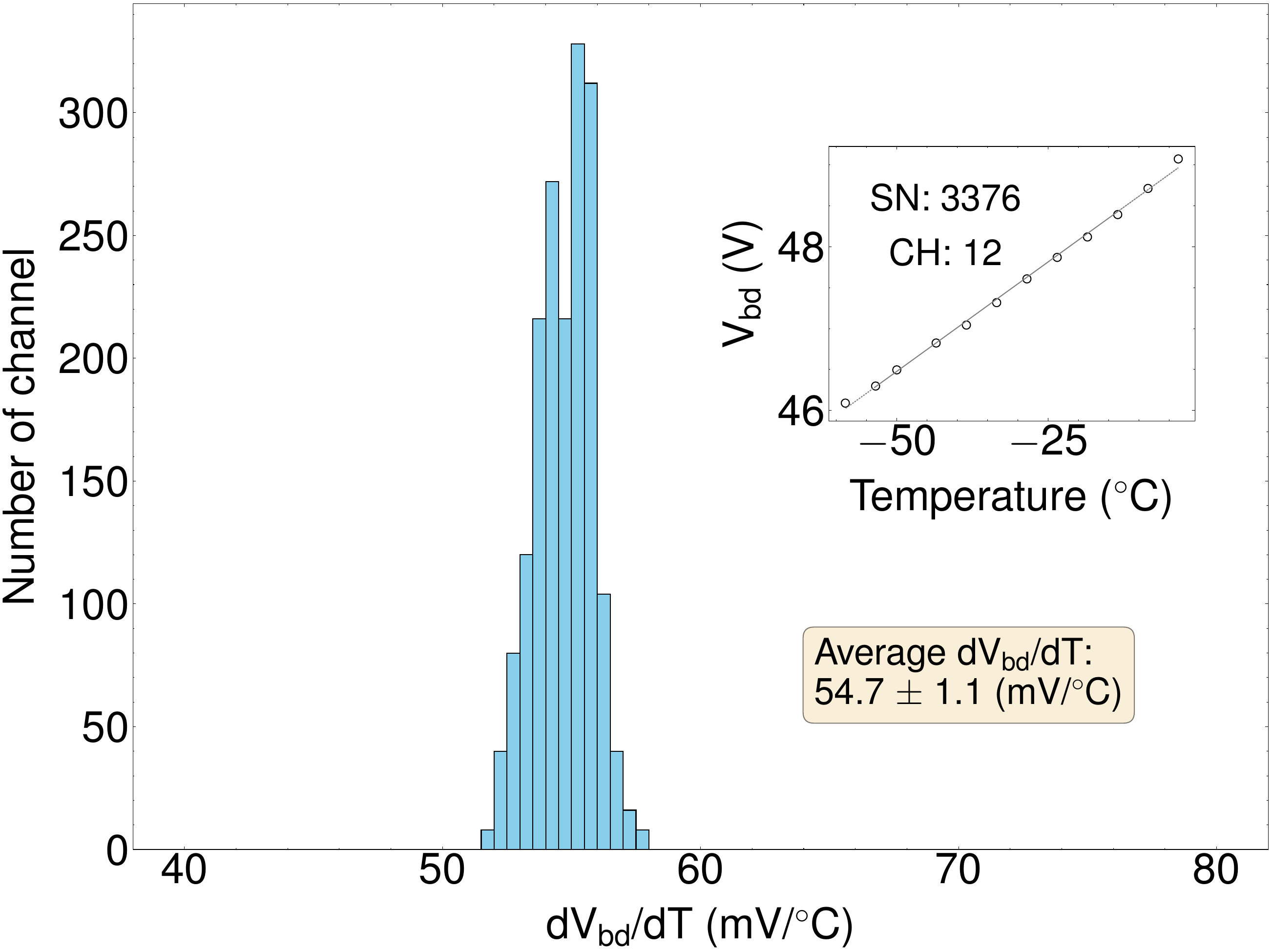}
    \caption{Breakdown voltage vs. temperature analysis for tested SiPMs: this figure illustrates the correlation between the breakdown voltage and operating temperature across the SiPMs evaluated.}
    \label{fig:vbd_temp}
\end{figure}

As described in the methods section of the work \cite{Rybnikov_2024}, we collected waveforms for each SiPM at six preset over-voltages under LED illumination. To determine the single photo-electron (SPE) charge at a given over-voltage, we fitted the charge spectrum—derived from the waveforms within the LED illumination range—with Gaussian peaks. The difference between these peaks provided the SPE charge. Subsequently, the breakdown voltage for each SiPM was determined through a linear regression analysis of SPE charge as a function of the applied voltage. The breakdown voltage is indicated by the x-intercept, where the charge gain equals zero.

Figure \ref{fig:vbd_temp} shows the correlation between the breakdown voltage and temperature for the tested SiPM channels, demonstrating that an increase in temperature corresponds to an elevation in breakdown voltage. The observed trend is quantitatively described by the slope of the linear regression, commonly referred to as the temperature coefficient of the breakdown voltage. For the selected SiPM channels in the test, this coefficient gives an average value of 54.7 $\pm$ 1.1 mV/$^{\circ}$C with the consideration of non-uniformity. This result is fairly consistent with 54 mV/$^{\circ}$C reported by Hamamatsu. 

\paragraph{Dark Count Rate}

The DCR is determined through waveform analysis, wherein a specific threshold is applied to identify DCR pulses. The DCR for the Hamamatsu SiPM S16088 is initially computed for each channel, and then normalized by the channel's area (144 mm$^2$), resulting in a measurement expressed in Hz/mm$^2$.
The analysis of the DCR as a function of temperature reveals critical insights into the behavior of our photodetector under varying thermal conditions. 

In this study, we employ a model analogous to that presented by Collazuol et al. (2011) \cite{COLLAZUOL2011389} and Aymeric et al. (2020) \cite{panglosse2020dark},  the relationship between the DCR and temperature, incorporating the parameter of the energy gap $E_g$, which is formulated as 
\begin{equation}
    \mathrm{DCR}(T) = A_\mathrm{amp} \cdot T^{3/2} \cdot \exp( - E_g / k T)
    \label{eq:dcr-slope-temp}
\end{equation}
In this equation, \(T\) represents the temperature in Kelvin, \(k\) is the Boltzmann constant, and \(E_g\) denotes the energy gap parameter, an intrinsic property that reflects the material's band structure. The coefficient \(A_\mathrm{amp}\) is a device-specific parameter that scales the DCR amplitude. 

\begin{figure}
    \centering
     \begin{subfigure}[b]{0.45\linewidth}
        \includegraphics[width=\linewidth]{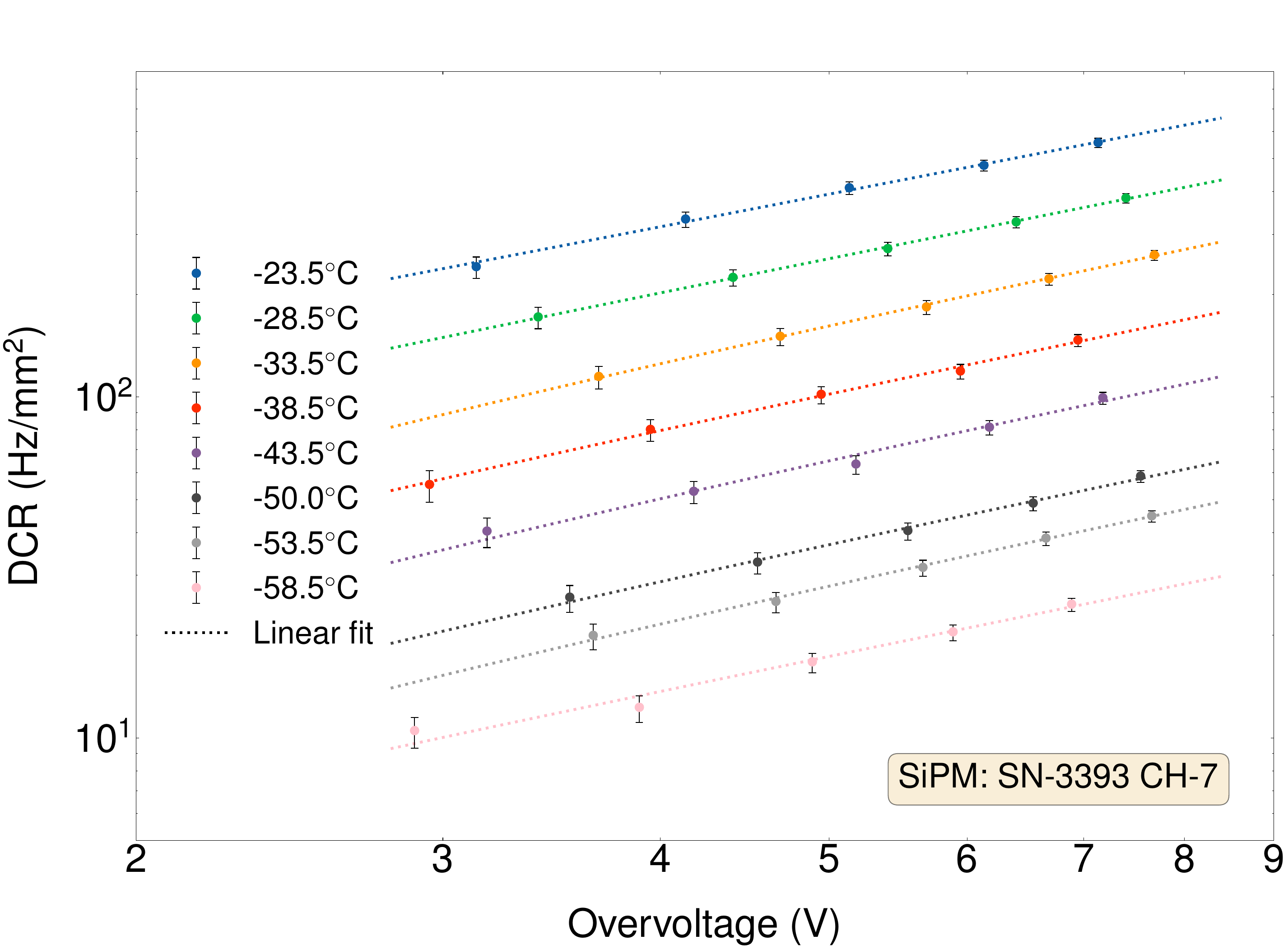}
        \caption{}
        \label{fig:dcr-vs-ov}
    \end{subfigure}
     \begin{subfigure}[b]{0.45\linewidth}
        \includegraphics[width=\linewidth]{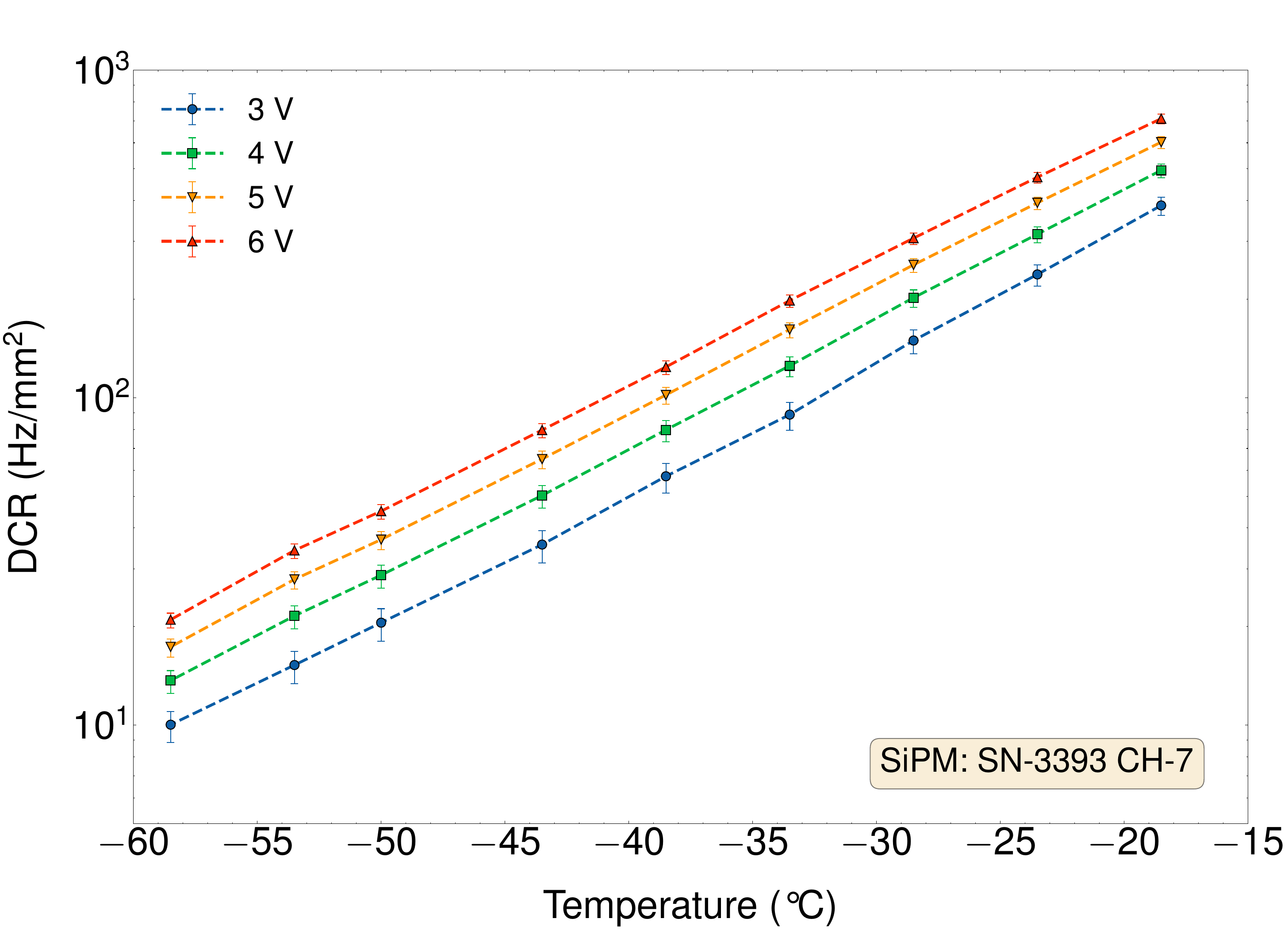}
        \caption{}
        \label{fig:dcr-vs-temp}
    \end{subfigure}
    
    \caption{(a) DCR versus overvoltage across various temperatures for a SiPM (b) Temperature dependence of DCR under varied overvoltages}

    \label{fig:dcr_ov_temp}
\end{figure}

\begin{figure}[htp]
    \centering
    \begin{subfigure}[b]{0.45\linewidth}
        \includegraphics[width=\linewidth]{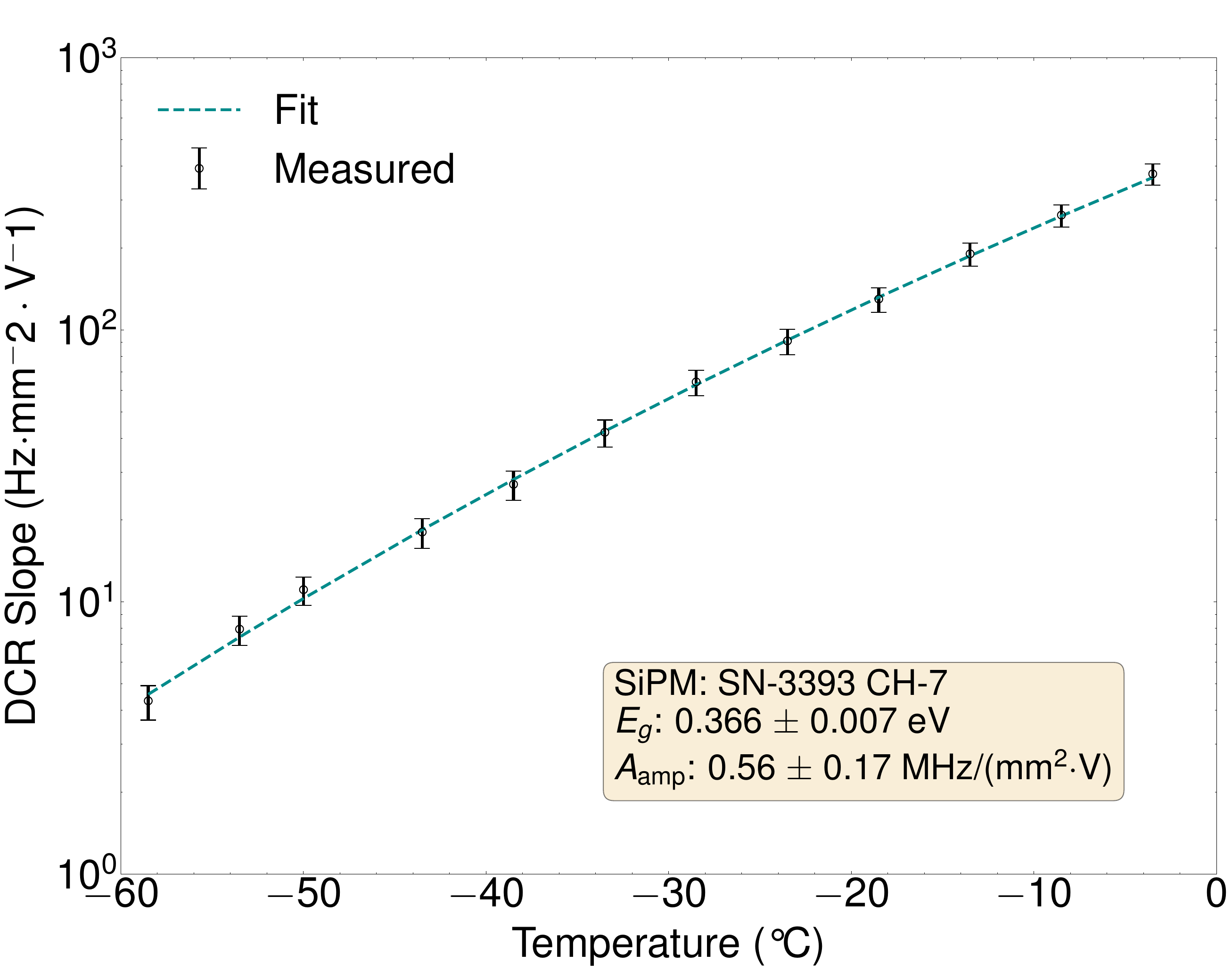}
        \caption{}
        \label{fig:dcr-slope}
    \end{subfigure}
    \begin{subfigure}[b]{0.45\linewidth}
        \includegraphics[width=\linewidth]{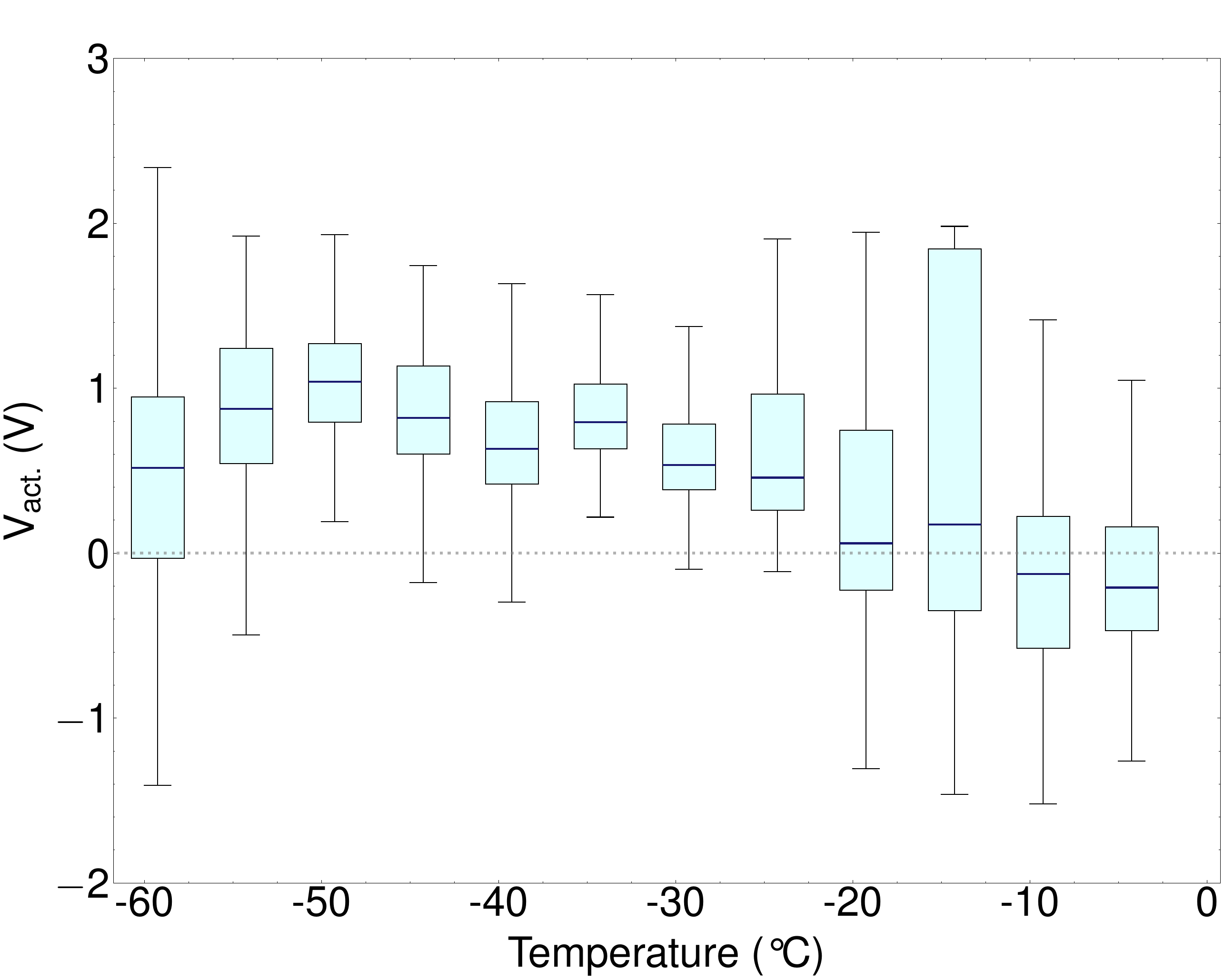}
        \caption{}
        \label{fig:dcr-intercept}
    \end{subfigure}
    \caption{(a) DCR v.s. overvoltage slope $\mathcal{S}(T)$ of the SiPM as a function of the temperature (left) (b) Baseline overvoltage $V_\mathrm{act}$ for DCR activation of the tested SiPMs (right)}
    \label{fig:dcr-temp}
\end{figure}

Results indicate that the DCR exhibits a good, albeit imperfect, linear relationship with overvoltage, as illustrated in Figure\,\ref{fig:dcr-vs-ov} with a specific SiPM across varying temperatures. Additionally, Figure\,\ref{fig:dcr_ov_temp} presents the DCR as a function of temperature for different overvoltages. Aligning with our linear approach to modeling DCR as a function of overvoltage, we prioritize presenting the slope $\mathcal{S}(T)$, defined as the rate of DCR alteration per volt derived from the linear fit, over direct DCR values. This strategy ensures consistency in our discussions and analysis. Figure \ref{fig:dcr-slope} displays $\mathcal{S}(T)$ across a temperature scan ranging from -60$^{\circ}$C to -20$^{\circ}$C, well fit by the relation from Equation \ref{eq:dcr-slope-temp}. This graph demonstrates how the DCR slope, derived from linearly fitting the DCR against overvoltage, varies with temperature for the selected SiPM. The analysis takes into account the non-zero activation voltage ($V_\mathrm{act}$) at specific temperatures, adjusting the DCR-temperature relation by considering a 1 V overvoltage increment to account for temperature dependence of $V_\mathrm{act}$, based on Equation \ref{eq:dcr-slope-temp}. It is important to note that due to the pile-up effect observed when calculating DCR at higher temperatures (>$-15 ^\circ$C) and higher overvoltages (>6V), the fit in the subsequent analysis does not include data affected by this effect. These data points are presented for illustrative purposes only.

It is important to note the potential temperature dependence of the energy gap, $E_g$. If $E_g$ varies with temperature such that $E'_g = a + b\cdot T + c\cdot T^2$, this introduces a modification to the energy gap parameter. In this scenario, the temperature invariant term $a$, which is equal to $E_g$, represents the energy gap at a reference temperature. The term $b\cdot T$ merges into the amplitude parameter $A_{\mathrm{amp}}$ during the fitting process, while the term $c\cdot T^2$ is neglected due to the limitations in the precision of DCR measurement and the relatively negligible value of $c$. This approach accounts for the dynamic nature of the energy gap with varying temperatures.

\begin{figure}[htb]
    \centering
    \begin{subfigure}[b]{0.48\linewidth}
        \includegraphics[width=\linewidth]{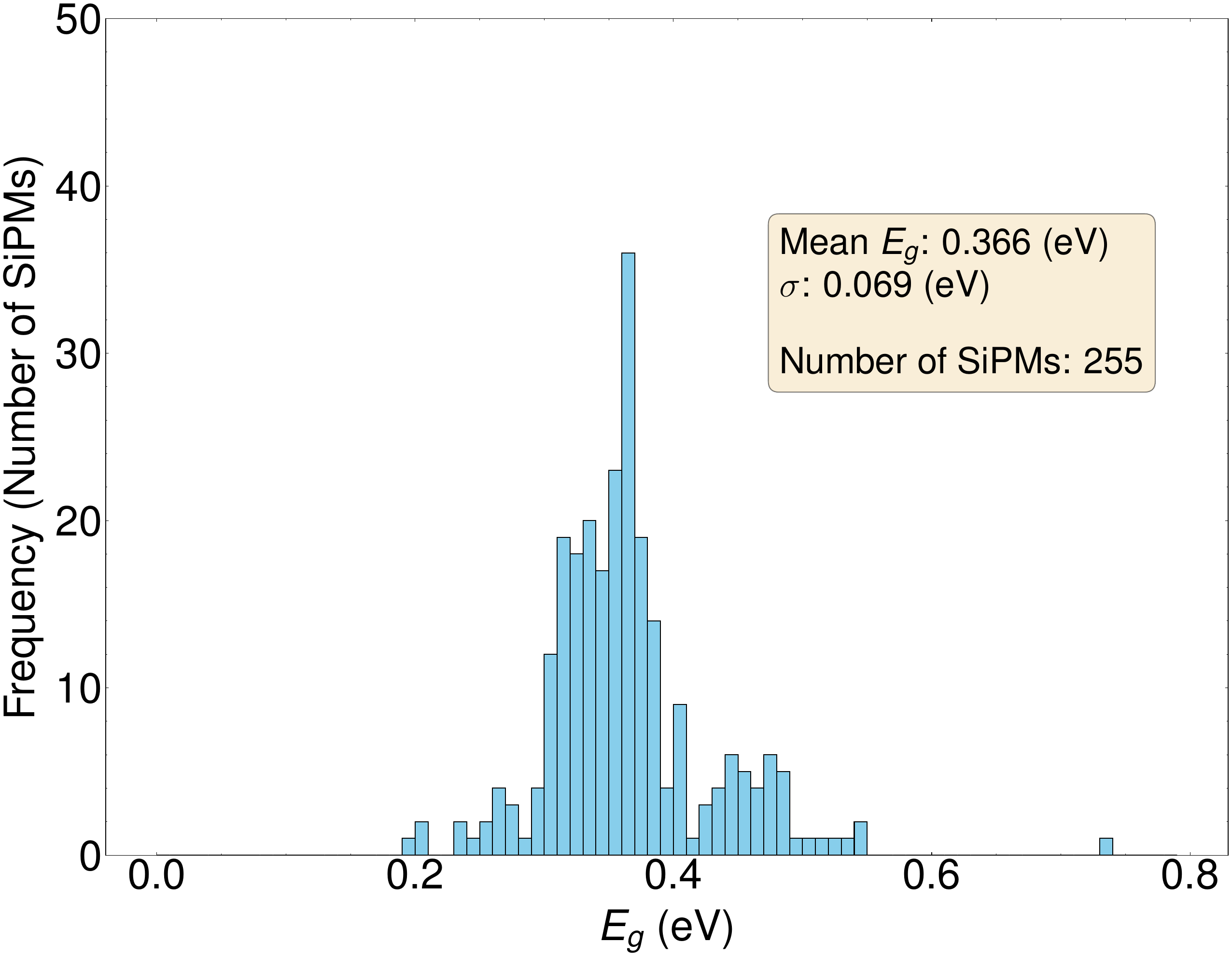}
        \caption{Energy Gap ($E_g$) Distribution}
        \label{fig:dcr-fit-a}
    \end{subfigure}
    \hfill
    \begin{subfigure}[b]{0.48\linewidth}
        \includegraphics[width=\linewidth]{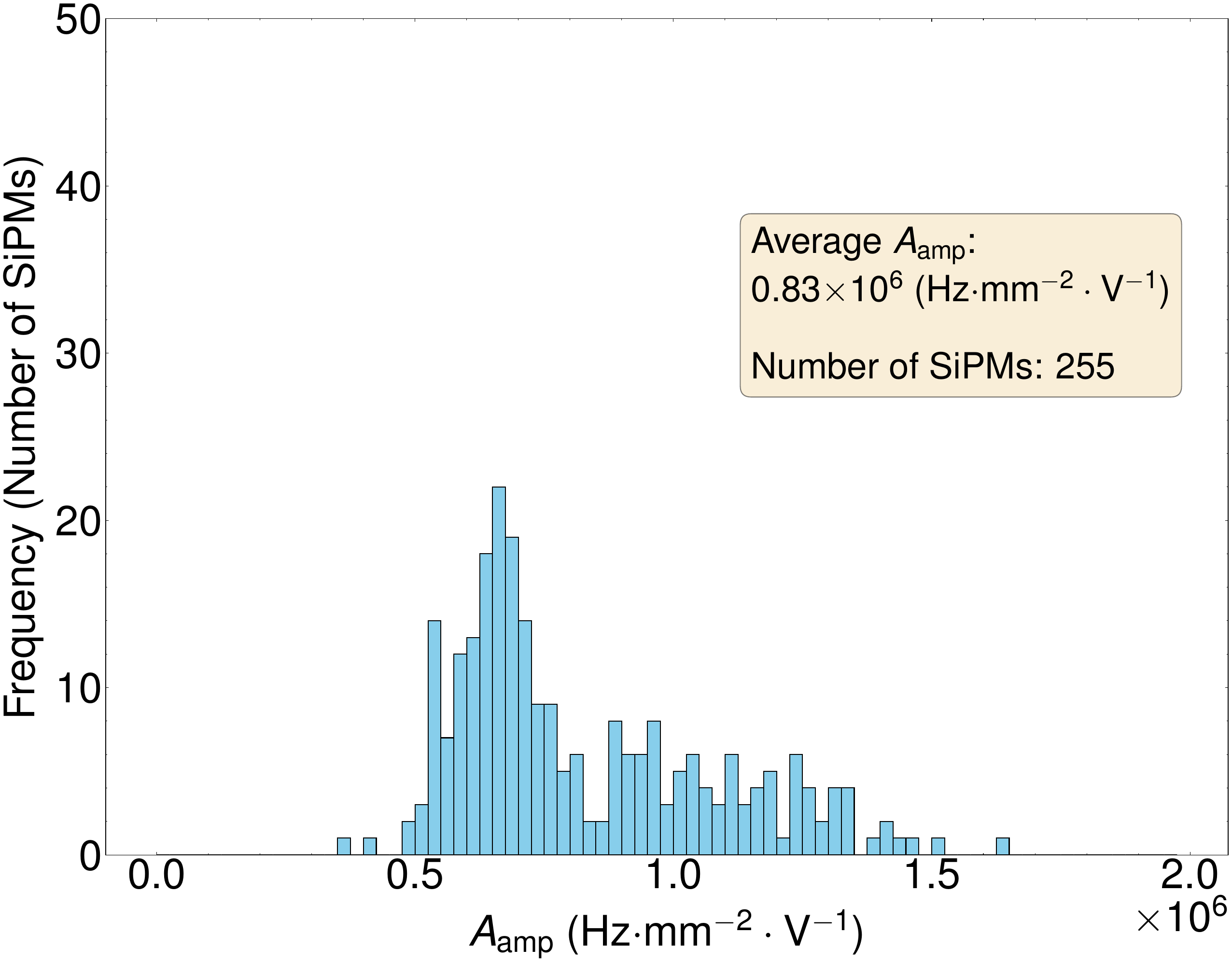}
        \caption{Amplitude ($A_{\mathrm{amp}}$) Distribution}
        \label{fig:dcr-fit-b}
    \end{subfigure}
    \caption{(a) Energy gap ($E_g$) distribution from the fit (b) Amplitude ($A_{\mathrm{amp}}$) distribution with fixed $E_g$}
    \label{fig:dcr-param-distribution}
\end{figure}

The data plotted in Figure \ref{fig:dcr-intercept} originates from the linear fit analysis of DCR as a function of the overvoltage wherein the starting overvoltage $V_\mathrm{act}$ is measured across different temperatures for 256 SiPMs. We used box-and-whisker plots to illustrate non-uniformity, including contributions from potential uncertainties of DCR and the linear model, with extended error bars. The box represents values ranging from the 25th to the 75th percentile while the central black line indicates the median value. The whiskers (the lines extending from the box on both sides) extend to 1.5 times the interquartile range (IQR). It's noted that precise determination of $V_\mathrm{act}$ from room temperature burn-in data is challenging; and we use assume that a negative overvoltage (biasing voltage below the breakdown voltage) should not yield a non-zero DCR; Therefore, in subsequent discussions of room temperature results, we approximate the DCR as proportional to the overvoltage with no intercept ($V_\mathrm{act} = 0$).

\begin{figure}
    \centering
    \includegraphics[width=0.5\linewidth]{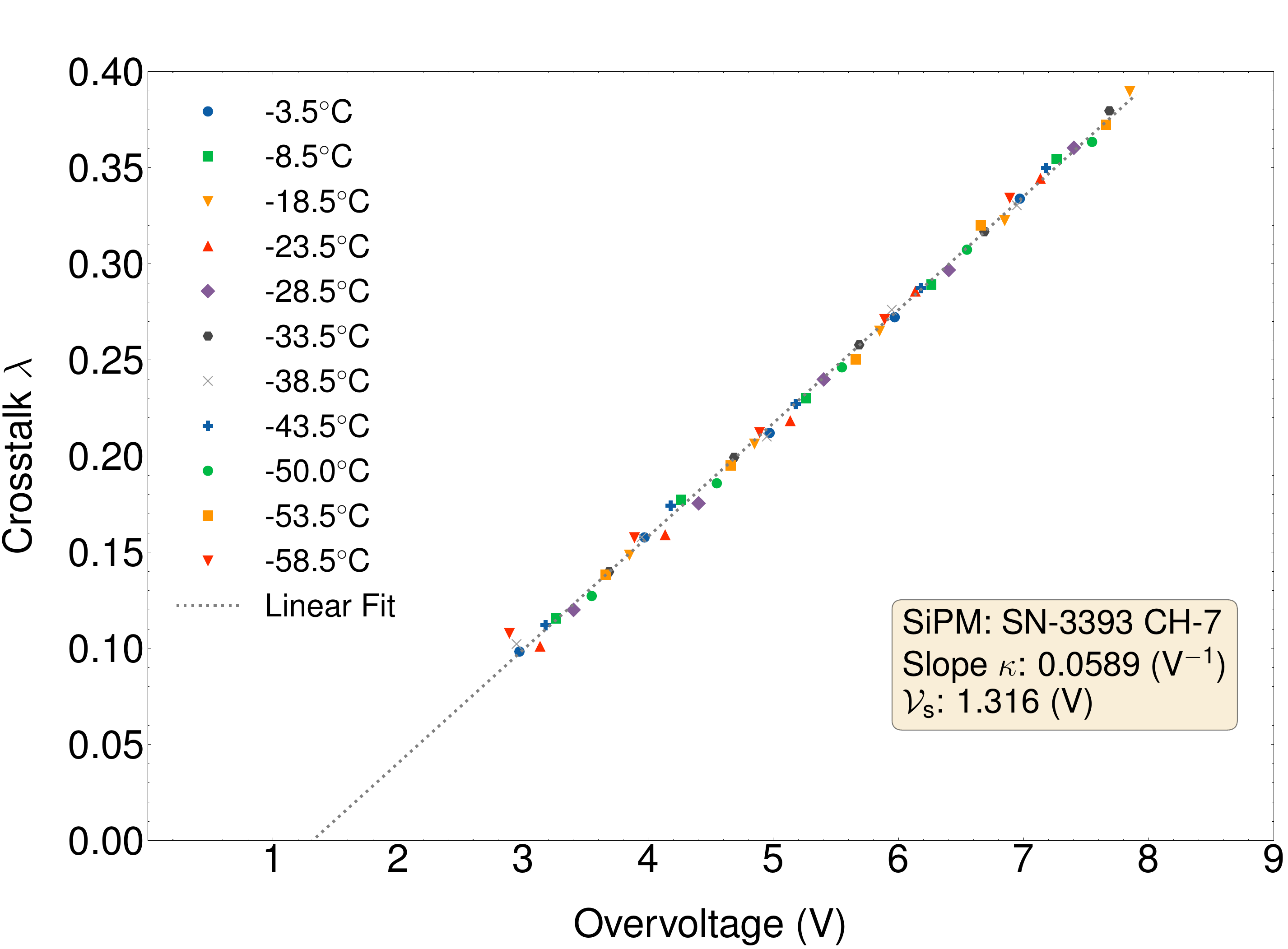}
    \caption{Variation of crosstalk parameter $\lambda$ with overvoltage in a selected SiPM at different temperatures}
    \label{fig:crosstalk}
\end{figure}

Through the analysis of 256 channels across 16 SiPM tiles, we fitted our experimental data to Equation \ref{eq:dcr-slope-temp} with two free parameters: $E_g$ and $A_{\mathrm{amp}}$. The distribution of $E_g$, as depicted in Figure \ref{fig:dcr-fit-a}, has an average value of $0.37\pm 0.07$ eV, which is consistent with the value of about 0.36 eV reported in \cite{COLLAZUOL2011389}. Figure \ref{fig:dcr-fit-b} reveals the spread of amplitude $A_{\mathrm{amp}}$ values across various SiPMs. To ensure consistency and focus on the amplitude's variation, the energy gap was fixed at 0.366 eV during the fit. It is important to recognize that such $E_g$ may differ across devices due to variations in doping concentrations or the presence of defects. Determining $E_g$ for each device, particularly under conditions of low temperatures, presents significant challenges. Therefore, we adopt a statistical approach to utilize $E_g$ values for describing the dark current, $I_\mathrm{dark}$, which will be elaborated upon subsequently.
\paragraph{Crosstalk}
Figure \ref{fig:crosstalk} shows how the crosstalk parameter, $\lambda$, varies with overvoltage at different temperatures for a specific SiPM. This parameter is derived by describing the number of photon-electrons of pulses with a statistical model discussed in \cite{VINOGRADOV2012247}. It is expected that $\lambda$ demonstrates a linear correlation with overvoltage over a specific range and remains consistent across various temperatures. We conduct linear fits for all 256 SiPMs and obtain the average values of $\kappa = 0.0604$ V$^{-1}$ and $\mathcal{V}_\mathrm{s} = 1.264$ V, which are adopted to facilitate the description of the dark current in the following discussions. We made this approximation for two reasons: 1) the variations of $\kappa$ and $\mathcal{V}_\mathrm{s}$ among different devices are small, and 2) it is not feasible to measure these values for all 64,000 SiPMs in over 4,000 SiPM tiles, given that we only obtained values for 256 SiPMs.
\subsection{Interpretation of the Burn-in Test Data}
We examined over 4,000 SiPM tiles in the burn-in test, periodically recording variations in current and temperature. 

\paragraph{Dark Current Fit}
In our study, we describe the dark current by Equation \ref{eq:est-current} treating the breakdown voltage at the reference temperature ($24^{\circ}$C) and the total current amplitude as two unconstrained parameters.

\begin{figure}[htb]
    \centering
    \begin{subfigure}[b]{0.48\linewidth}
        \includegraphics[width=\linewidth]{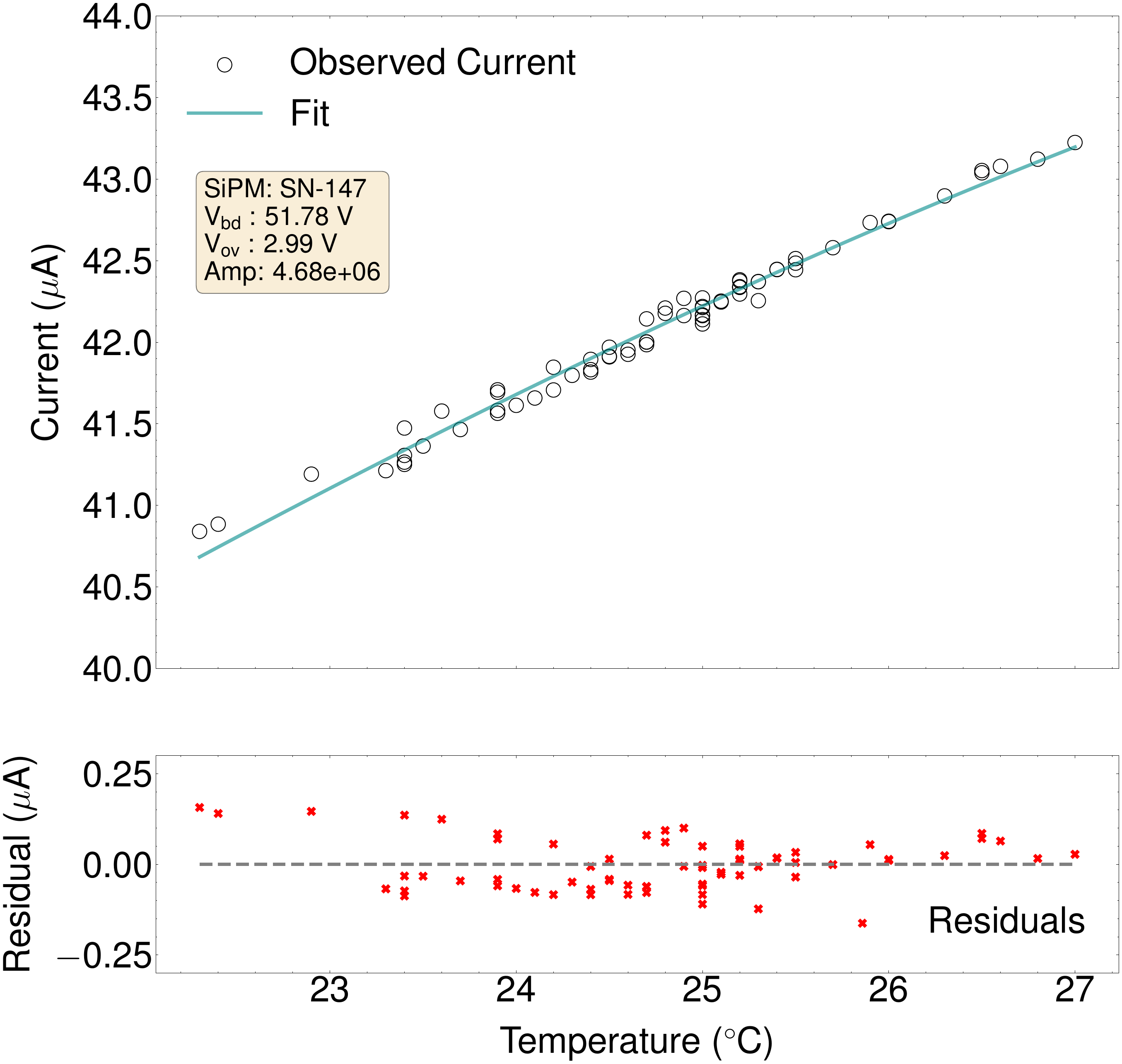}
        \caption{}
        \label{fig:prefit-temp}
    \end{subfigure}
    \hfill
    \begin{subfigure}[b]{0.48\linewidth}
        \includegraphics[width=\linewidth]{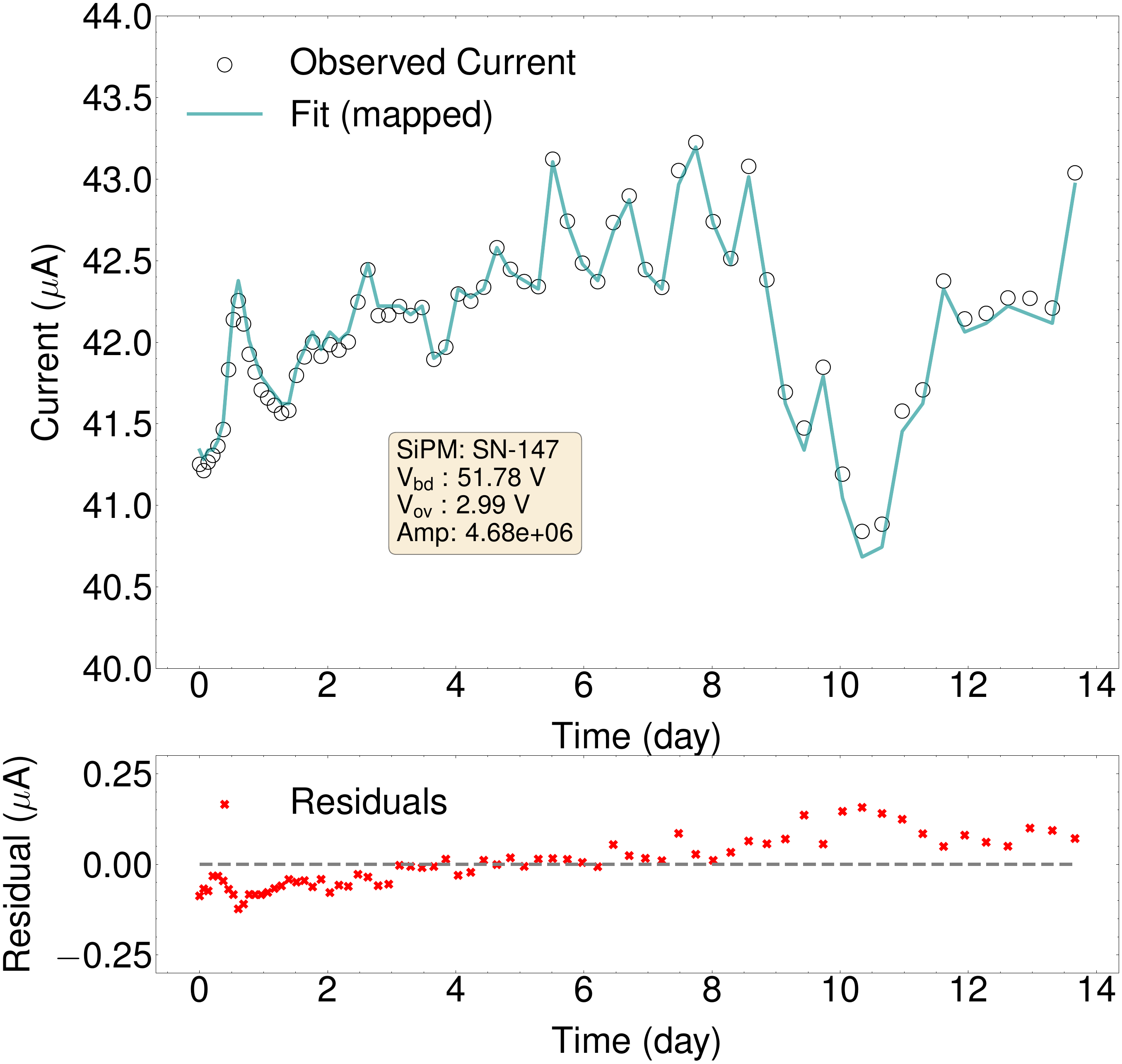}
        \caption{}
        \label{fig:prefit-time}
    \end{subfigure}
     \caption{Dark current model of temperature-only dependence: (a) Initial temperature fit showing how observed current values align with model predictions across different temperatures; (b) Corresponding translation of (a) where data and fit results are mapped to the recorded time.}
      \label{fig:prefit-burnin}
\end{figure}

Our model demonstrates a good fit to the burn-in data, accurately capturing the relationship between current and temperature, as illustrated in Figure\,\ref{fig:prefit-temp} where the data points are well-aligned with temperature variations. Recognizing that temperature can fluctuate over time, we extended our analysis to examine the stability. This was achieved by mapping temperature $T(t)$ readings over time. Notably, we observed a slight increase in dark current as shown in the residuals of Figure\,\ref{fig:prefit-time}.

We consider that heat accumulation contributed to the increase in dark current over time, as the recorded temperature may not accurately reflect the actual temperature of the silicon sensors. Therefore, we introduce a time-dependent term to the dark current model as a temperature correction:
\begin{equation}
\label{eq:time-constant}
I_\mathrm{t}(t) = A_R \cdot \left(1 - \exp\left(-\frac{t}{\tau}\right)\right),
\end{equation}
and subsequently, we modify the dark current model to include this term:
\begin{equation}
\label{eq:model2}
    I'_\mathrm{dark}(T, t) = I_\mathrm{dark}(T) + I_\mathrm{t}(t),
\end{equation}
where \(A_R\) represents the amplitude of this effect, and \(\tau\) is the time constant associated with this process. Both \(A_R\) and \(\tau\) serve as free parameters and are determined through a combined fit of the model to the data.

\begin{figure}[htb]
    \centering
    \begin{subfigure}[b]{0.48\linewidth}
        \includegraphics[width=\linewidth]{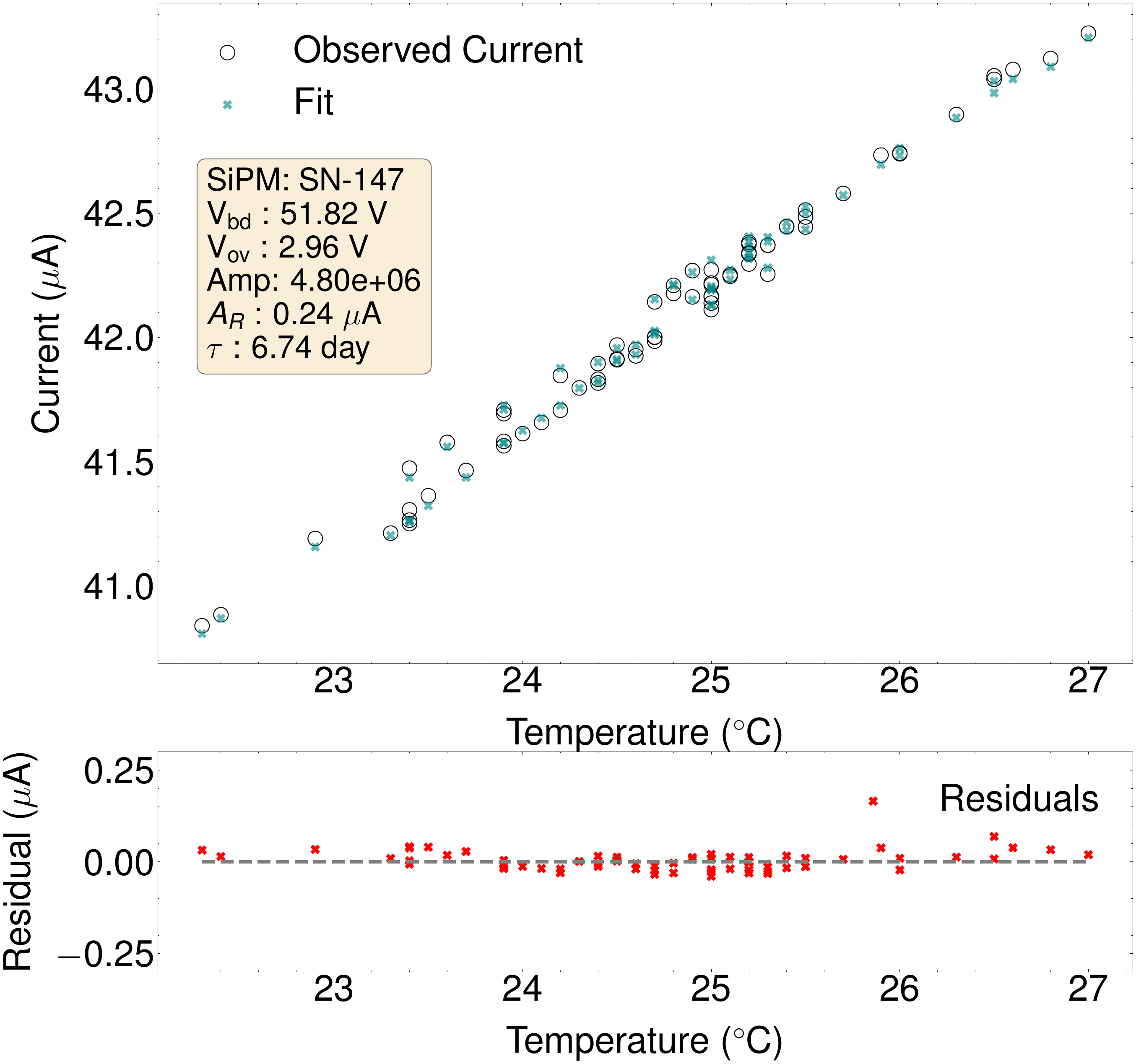}
        \caption{}
        \label{fig:timefit-temp}
    \end{subfigure}
    \hfill
    \begin{subfigure}[b]{0.48\linewidth}
        \includegraphics[width=\linewidth]{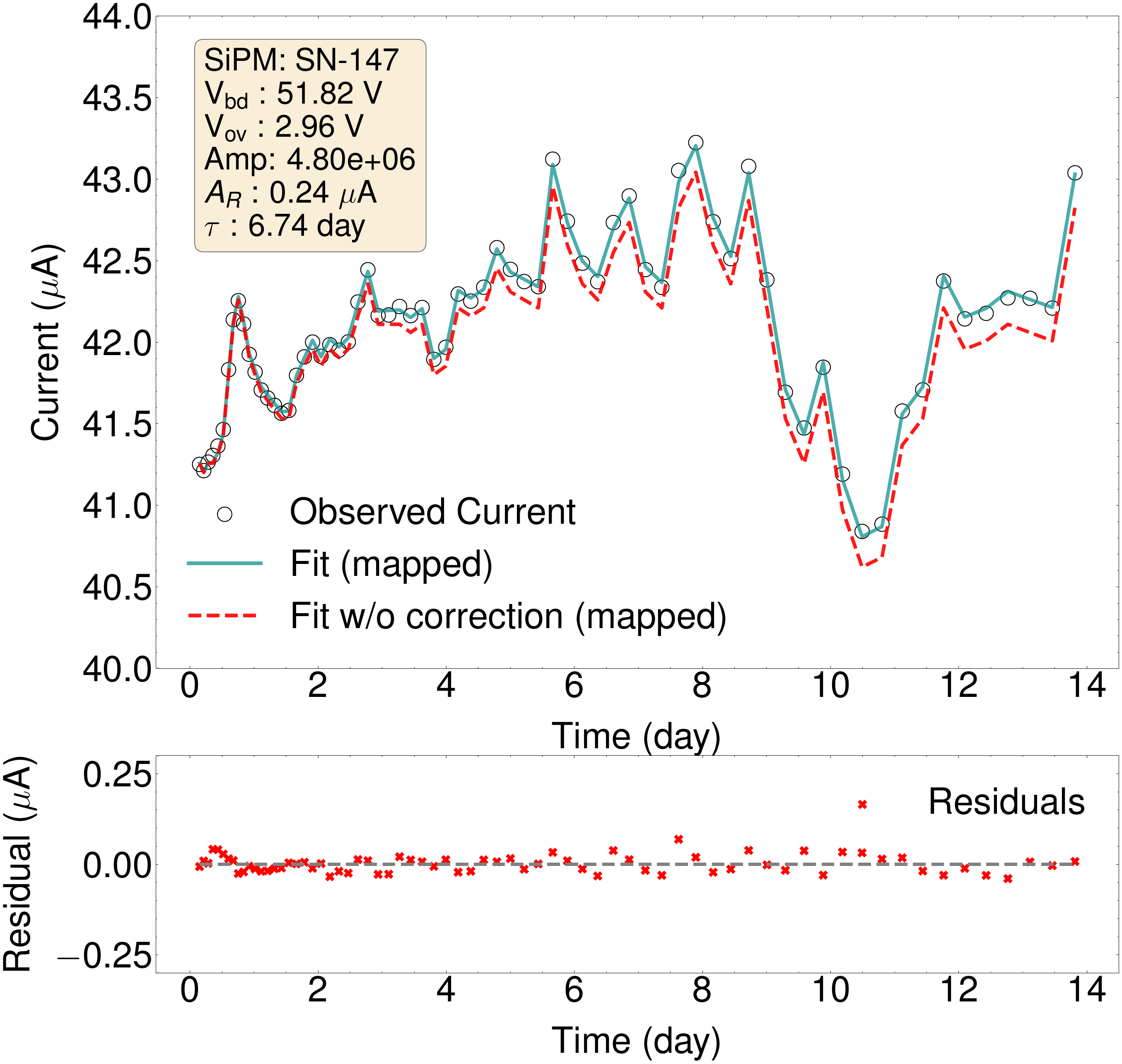}
        \caption{}
        \label{fig:timefit-time}
    \end{subfigure}
    \caption{Temperature-time-dependent model validation against burn-in data: (a) showcases the alignment between observed dark current values and those predicted by our temperature-time-dependent model across a range of temperatures, (b) depicts the dark current's time evolution, comparing measured values against those predicted after applying temperature correction}
    \label{fig:timefit-burnin}
\end{figure}
\begin{figure}[htb]
    \centering
    \begin{subfigure}[b]{0.48\linewidth}
        \includegraphics[width=\linewidth]{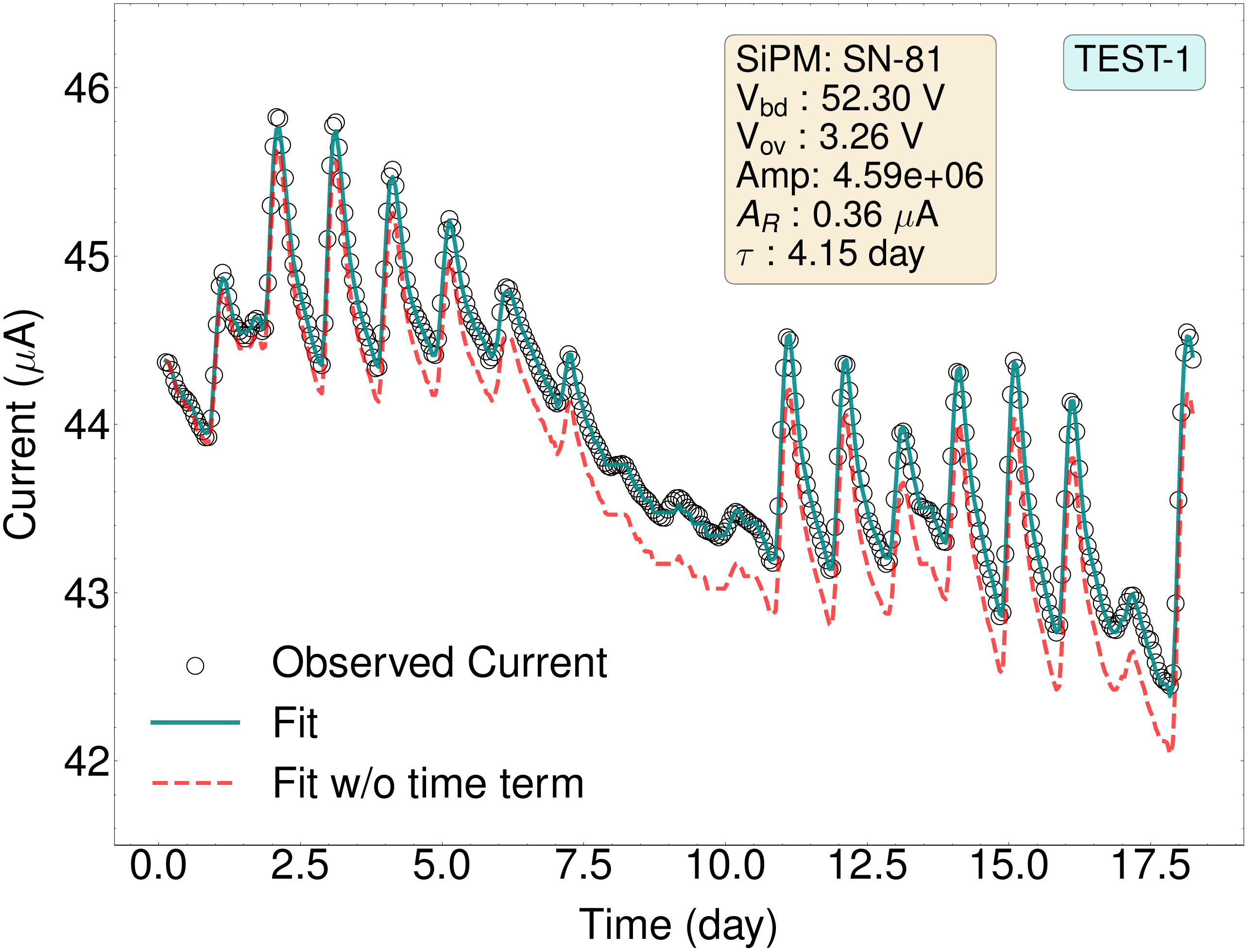}
        \caption{First burn-in test}
        \label{fig:twice-fit-test1}
    \end{subfigure}
    \hfill
    \begin{subfigure}[b]{0.48\linewidth}
        \includegraphics[width=\linewidth]{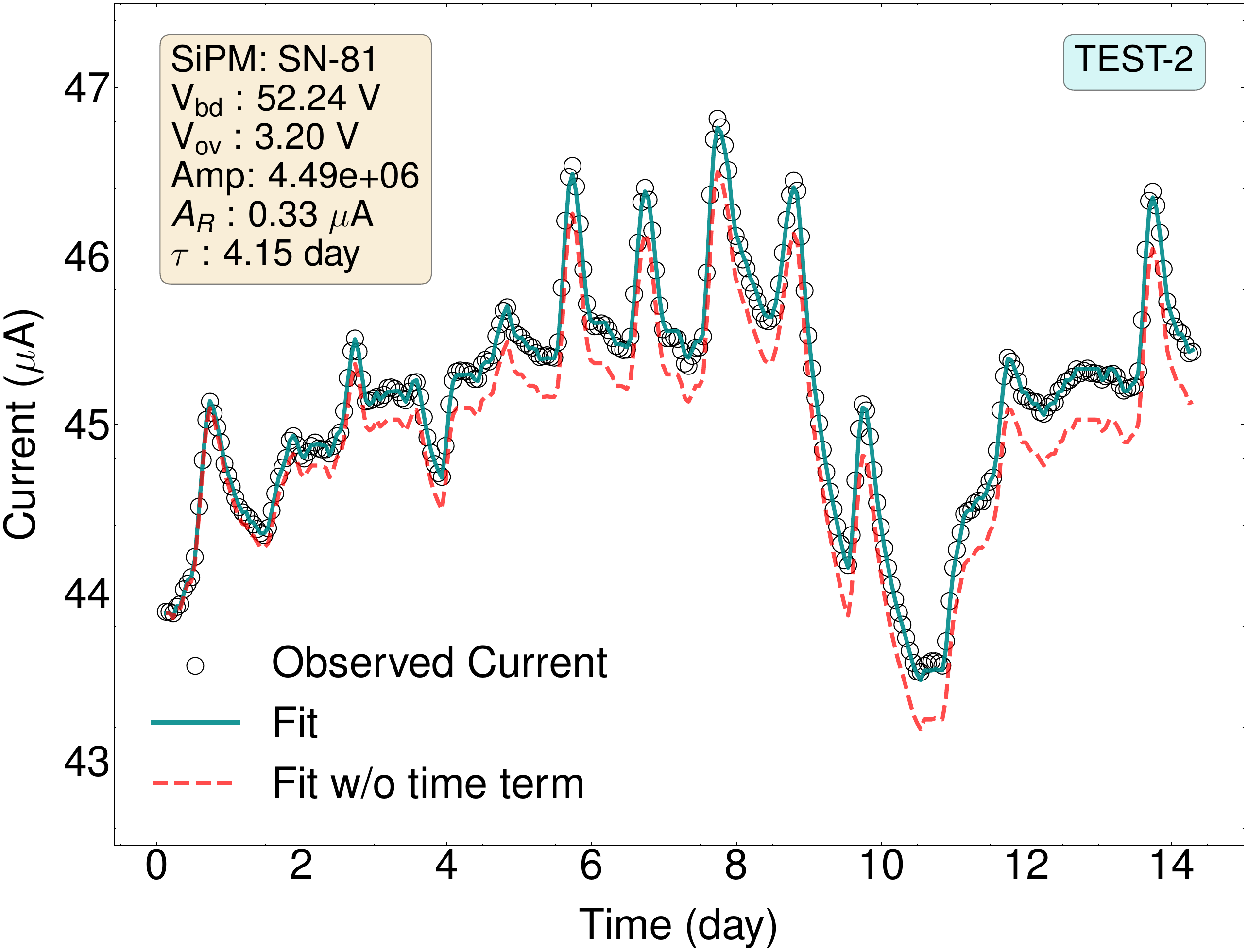}
        \caption{Second burn-in test}
        \label{fig:twice-fit-test2}
    \end{subfigure}
    \caption{Two distinct burn-in tests of the same SiPM}
    \label{fig:twice-burnin}
\end{figure}

After incorporating the temperature correction over time, the fit describes the measured data more precisely, as illustrated in Figure\,\ref{fig:timefit-temp} and \ref{fig:timefit-time}. The combined fit, incorporating both temperature and time variables, represents a two-dimensional analysis. Given the complexity of visualizing a 2D fit in a single plot, we present the results separately: one plot for temperature and another for time, to enhance clarity. Additionally, a one-dimensional fit for temperature alone is included in Figure\,\ref{fig:timefit-time}, shown as a red dashed line, for comparison after being mapped to the recorded time.
We also conducted a detailed examination on a subset of SiPMs that underwent the burn-in test on two separate occasions, as depicted in Figure\,\ref{fig:twice-burnin}. Each SiPM was initially subjected to the burn-in test, removed from the experimental setup, and subsequently reintroduced for a second test. The reiteration of the effect, with consistent levels of the parameters $A_R$ and $\tau$, suggests that what we observed is not attributable to permanent aging of the SiPM. Instead, it is more likely to reflect a transient, potential thermal inertia of the silicon sensors during the test.

It should be noted that Equation\,\ref{eq:time-constant} introduces two parameters, $A_R$ and $\tau$, to take into account empirically the drift of the real temperature of the device from its measurement. These parameters vary among different devices and change with different testing conditions, so they are not intended to describe the behavior of individual devices. Therefore, this term should be considered as a parameterization of systematic temperature variation, and it is introduced to reduce systematic errors.

\begin{figure}[htb]
    \centering
    \includegraphics[width=0.66\linewidth]{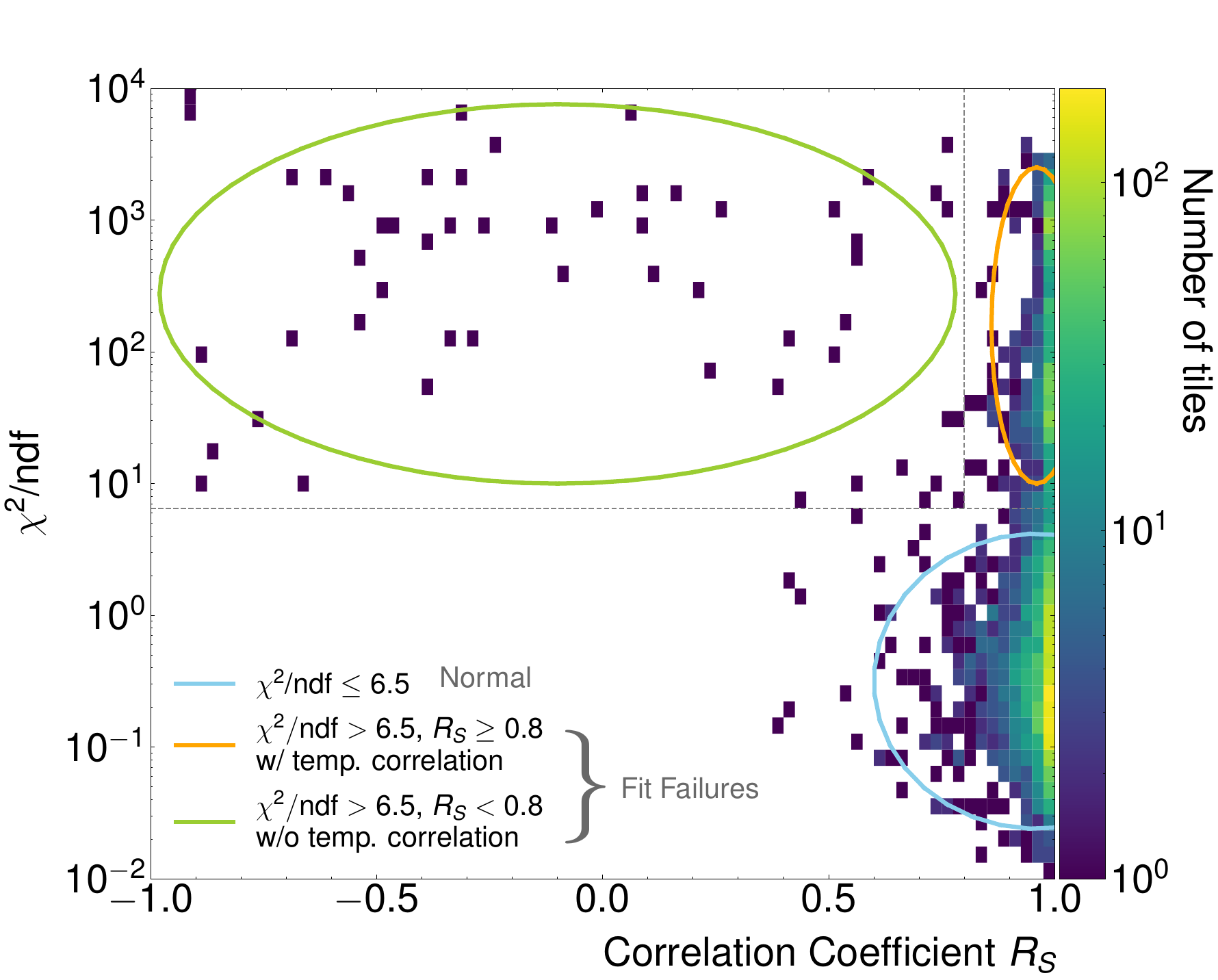}
    \caption{Distribution of normalized $\chi^2$ values and Spearman's rank correlation coefficients for SiPM tiles based on dark current modeling.}
    \label{fig:burn-in-chi2}
\end{figure}

We also introduce a goodness-of-fit measure, $\chi^2 = \sum (\hat{I}-I)^2 / (\sigma^2_x + \sigma^2_y)$, to quantitatively assess the agreement between our model's predictions and the actual observed data. In this context, $\sigma_x = I_\mathrm{dark}(T + T_{\mathrm{unc.}}) - I_\mathrm{dark}(T)$ accounts for the uncertainty in temperature measurements, where $T_{\text{unc.}} = 0.1$ °C reflects the precision of the temperature sensor and potential fluctuations. Meanwhile, $\sigma_y$ represents the standard error associated with the measured currents. In this formulation, $I$ denotes the observed dark currents, and $\hat{I}$ corresponds to the model's predictions. Figure~\ref{fig:burn-in-chi2} displays the normalized $\chi^2$ ($\chi^2$ divided by degrees of freedom) distribution across all SiPM tiles derived from the dark current modeling, alongside the Spearman rank correlation coefficients, which are calculated based solely on the temperature dependence of the dark current. We employ a $\chi^2$/ndf threshold of 6.5 and a correlation coefficient $R_S$ of 0.8 as criteria to distinguish between normal and abnormal SiPM tiles, as well as to identify fit failures. 

\begin{figure}[htb]
    \centering
    \begin{subfigure}[b]{0.48\linewidth}
        \includegraphics[width=\linewidth]{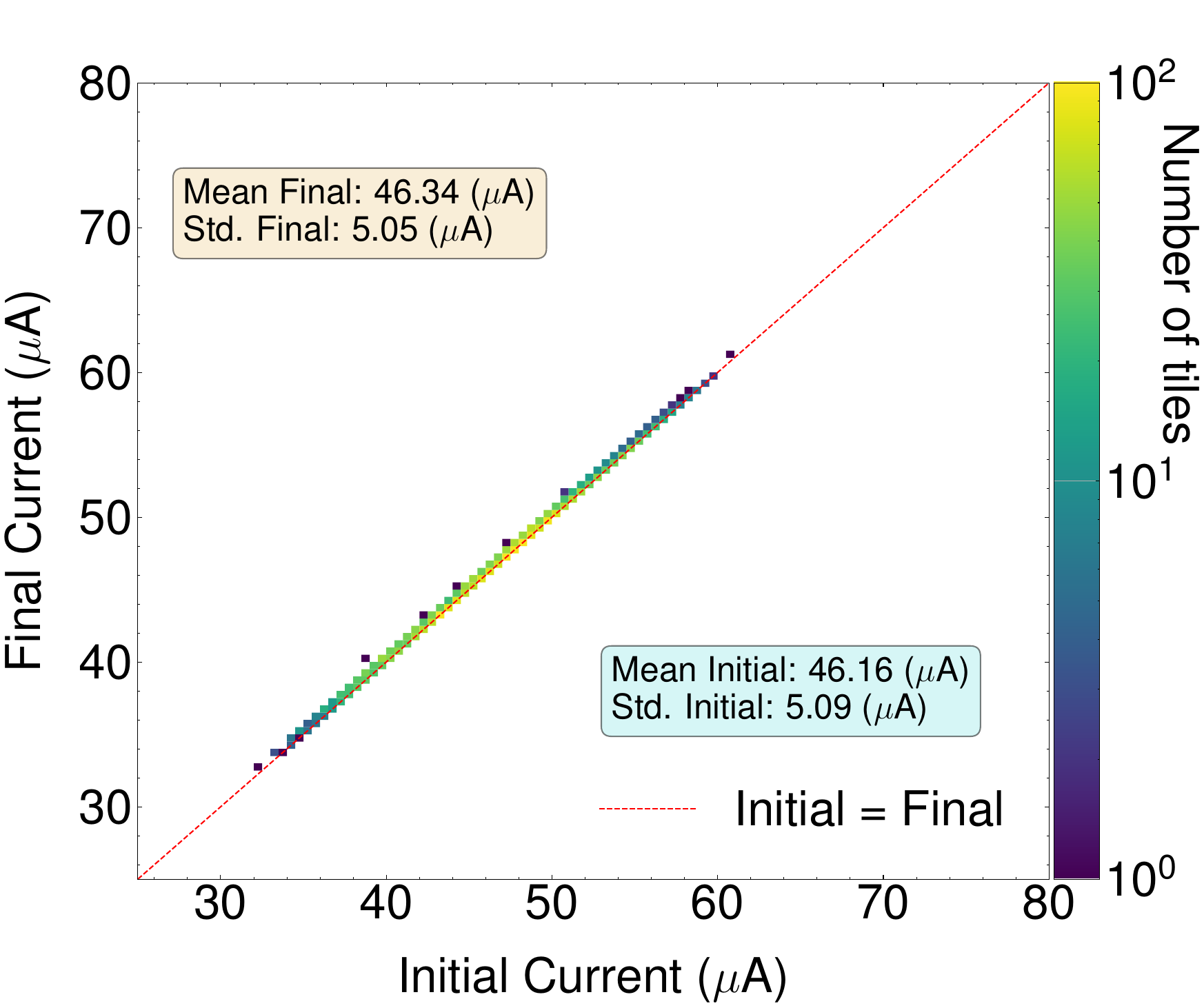}
        \caption{Normal SiPM tiles showing a tight distribution of dark currents ($\chi^2 /\mathrm{ndf}\leq 6.5$)}
        \label{fig:normal-current-dist}
    \end{subfigure}
    \hfill
    \begin{subfigure}[b]{0.48\linewidth}
        \includegraphics[width=\linewidth]{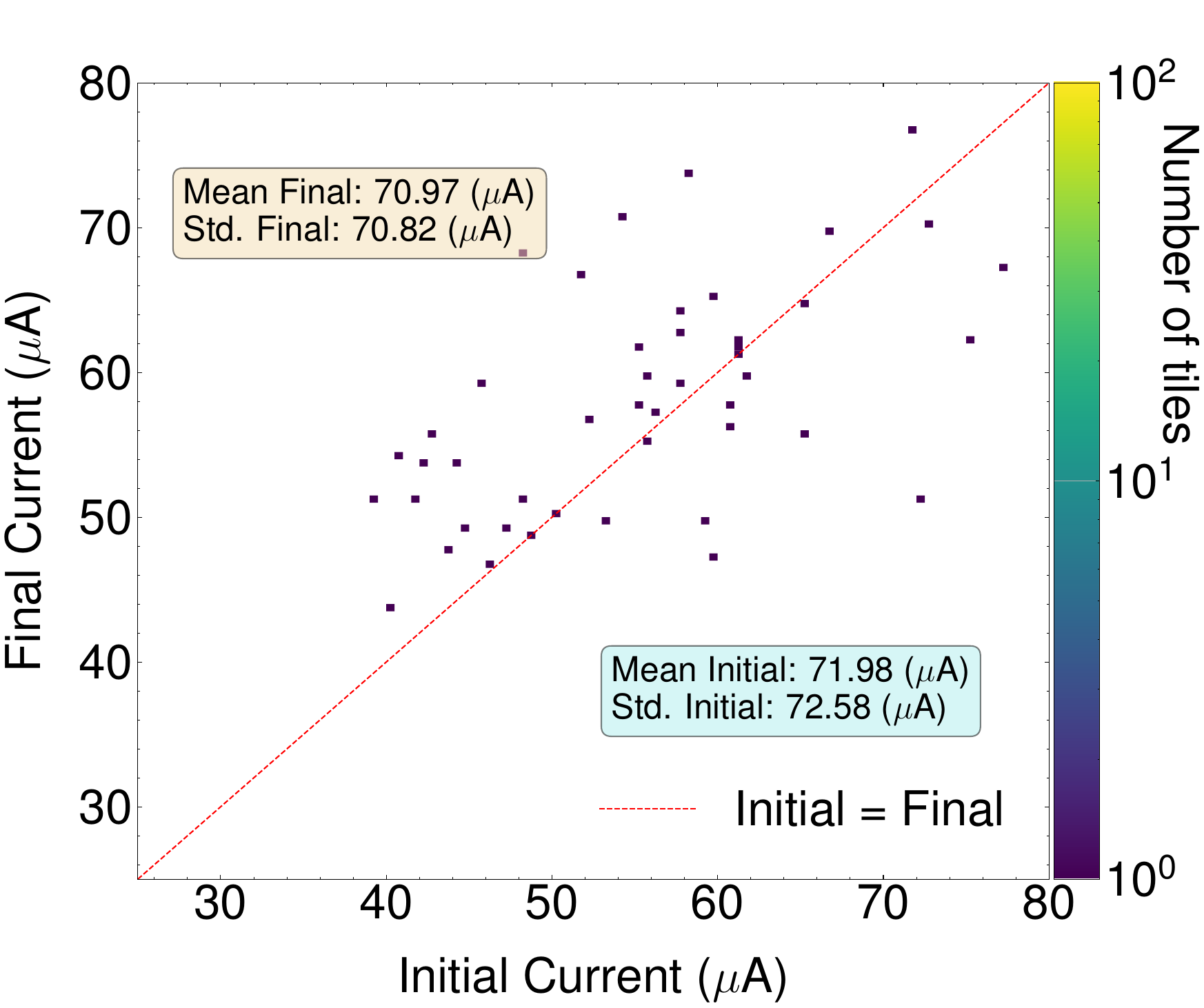}
        \caption{Abnormal SiPM tiles with a wide distribution of dark currents ($\chi^2/\mathrm{ndf} > 6.5$ $\cap$ $R_S < 0.8$ )}
        \label{fig:abnormal-current-dist}
    \end{subfigure}
    \caption{Comparison of dark current distributions for normal and abnormal SiPM tiles at a normalized average temperature.}
    \label{fig:current-dist}
\end{figure}

Following the application of these criteria, Figure~\ref{fig:current-dist} presents the distribution of dark currents for both normal and abnormal SiPM tiles. The dark current values are corrected to a reference temperature of 24$^{\circ}$C to facilitate direct comparison. Correction for normal SiPM tiles is applied using the fit model, whereas for the abnormal tiles, a linear temperature coefficient of current ($dI/dT$ = 0.4752 $\mu$A/$^{\circ}$C) derived from the normal tiles is used. This approach underscores the significant impact of thermal performance on dark current variability, with normal tiles displaying more consistent behavior than their abnormal counterparts. This analysis aids in identifying abnormal SiPM tiles that may require further examination.

Results of tested SiPM tiles are categorized by goodness-of-fit, correlation coefficient and dark current level in Table \ref{tab:cat-current}.
In the examination of burn-in test data, approximately 350 SiPM tiles failed ($\chi^2/\mathrm{ndf} > 6.5$) to meet our goodness-of-fit criteria, though they demonstrated a dark current and temperature correlation ($R_S \geq 0.8$). Most of these tiles were associated with a particular testing session, during which there was an abrupt rise in environmental temperature, disrupting the subtle thermal equilibrium. Judging from the dark current level, approximately 11 SiPM tiles were also distinguished as anomalies due to their high dark current ($\geq 100$ $\mu$A) behaviour, yet they still exhibited a consistent current-temperature correlation. These tiles have been marked for further investigation.

\begin{figure}[htb]
    \centering
        \begin{subfigure}[b]{0.45\linewidth}
        \includegraphics[width=\linewidth]{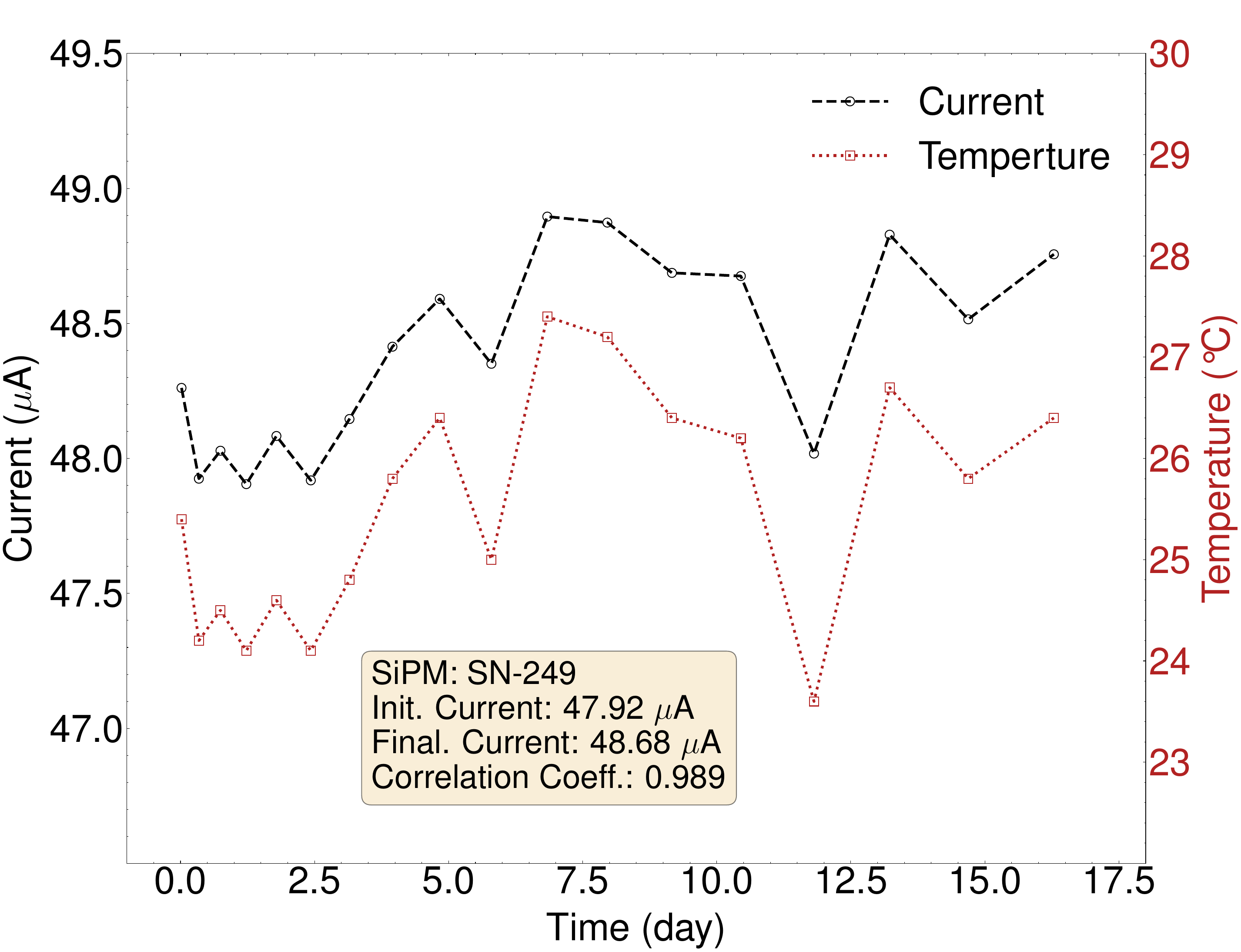}
        \caption{Normal}
        \label{fig:normal-current-time}
    \end{subfigure}
    \hfill
   \begin{subfigure}[b]{0.45\linewidth}
        \includegraphics[width=\linewidth]{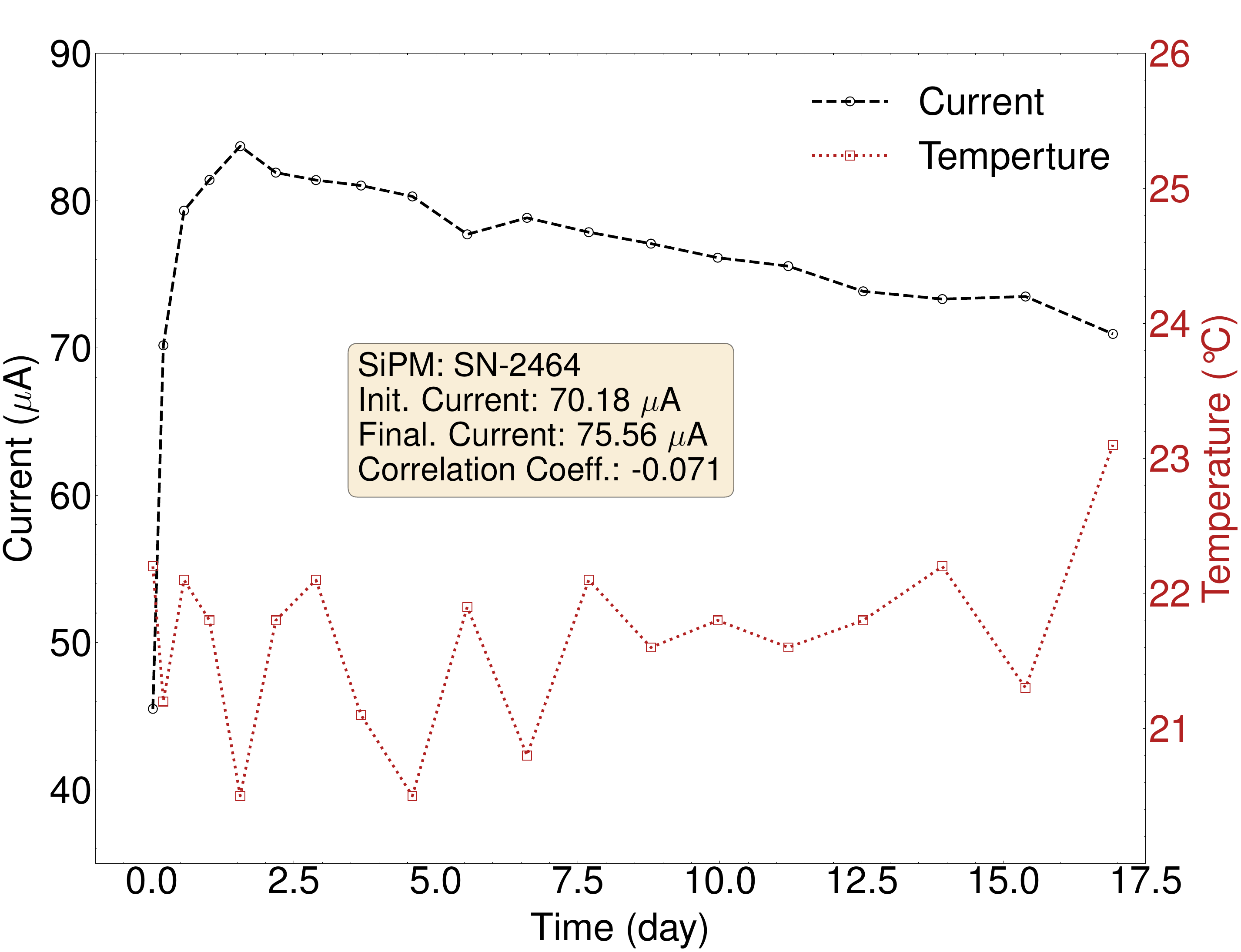}
        \caption{Abnormal}
        \label{fig:abnormal-current-time}
    \end{subfigure}

    \caption{Comparative illustration of temperature and dark current variations over time for normal and abnormal SiPM tiles.}
    \label{fig:current-temp-vs-time}
\end{figure}
The current and temperature data along the timeline for two distinct SiPMs, one exhibiting normal behavior (good current and temperature correlation) and one exhibiting abnormal behavior (poor current and temperature correlation), are shown in Figure\,\ref{fig:normal-current-time} and \ref{fig:abnormal-current-time}.

\begin{figure}
    \centering
    \includegraphics[width=0.75\linewidth]{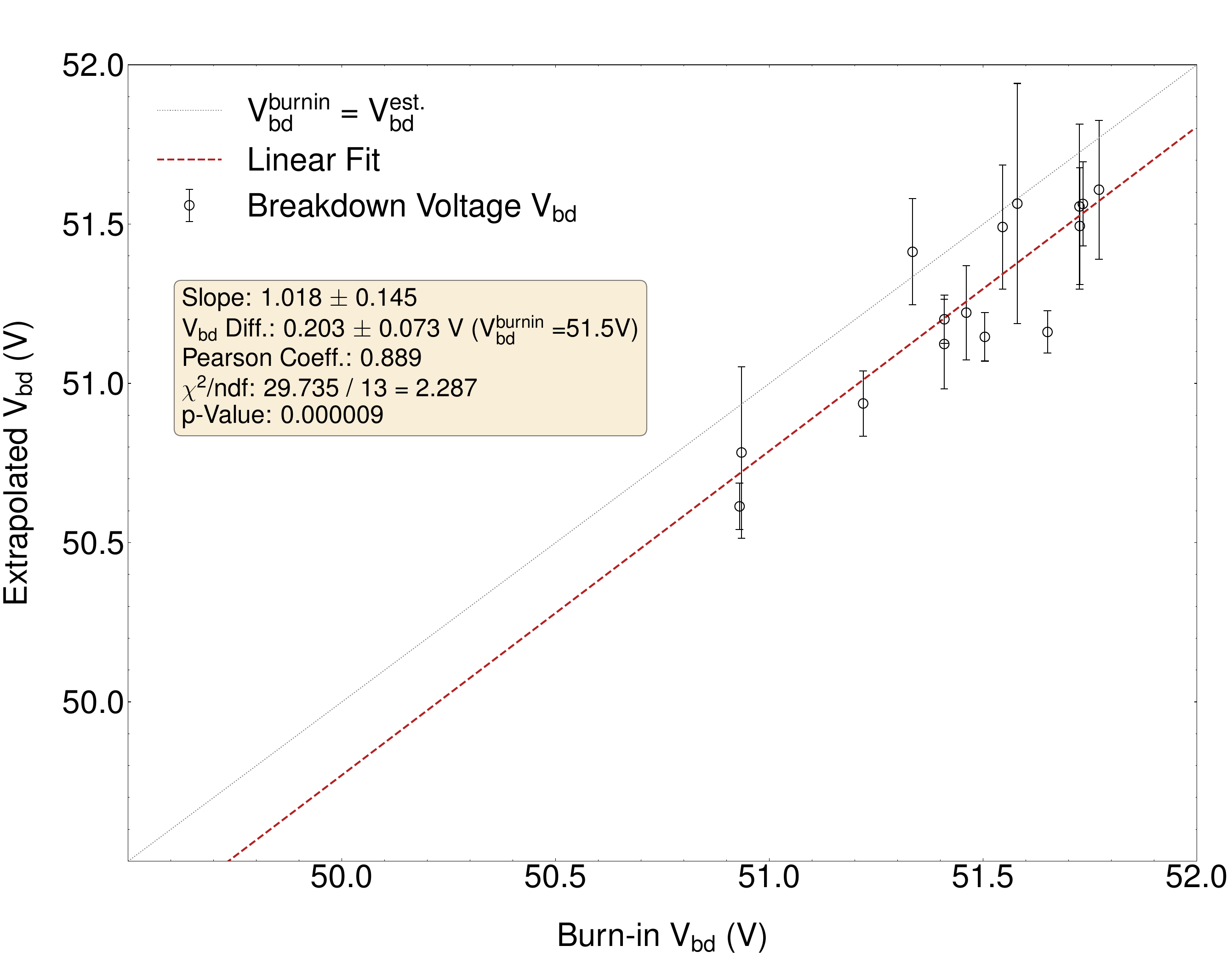}
    \caption{Linear comparison of breakdown voltage (V$_\mathrm{bd}$) from burn-in data and cryogenic test extrapolation}
    \label{fig:compare-vbd}
\end{figure}

We also conducted a linear comparison between the breakdown voltage, \(\mathrm{V}^\mathrm{burnin}_\mathrm{bd}\), derived from burn-in data fitting, and the estimated breakdown voltage, \(\mathrm{V}^\mathrm{est}_\mathrm{bd}\), extrapolated from cryogenic test results. This analysis includes 16 SiPM tiles, excluding one due to abnormal behavior. As shown in Figure\,\ref{fig:compare-vbd}, linear fitting reveals a strong correlation between \(\mathrm{V}^\mathrm{burnin}_\mathrm{bd}\) and \(\mathrm{V}^\mathrm{est}_\mathrm{bd}\), evidenced by a Pearson correlation coefficient of 0.889. The difference in breakdown voltage is 0.203 \(\pm\) 0.073 V at a baseline \(\mathrm{V}^\mathrm{burnin}_\mathrm{bd}\) of 51.5 V.

\begin{figure}
    \centering
    \includegraphics[width=0.66\linewidth]{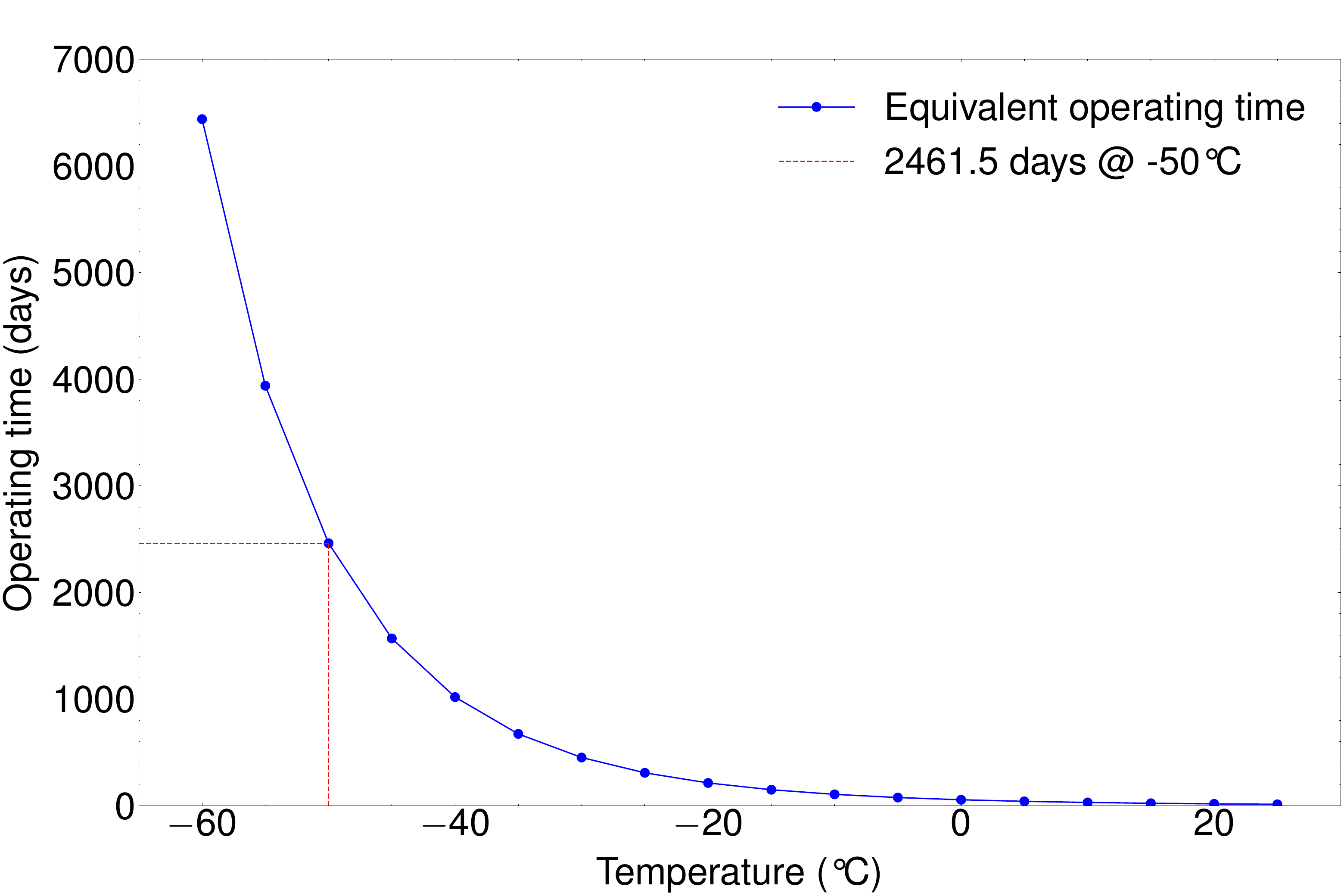}
    \caption{Equivalent operating time at various temperatures for a two-week burn-in test at room temperature (24°C).}
    \label{fig:equivalent-time}
\end{figure}

Finally, assuming constant charge endurance and negligible pixel gain variation with temperature, we calculate the equivalent operating time using Equation\,\ref{eq:dcr-slope-temp} with $E_g =$ 0.366 eV, as shown in Figure\,\ref{fig:equivalent-time}. The equivalent operating time at -50$^\circ$C for a two-week burn-in test at 24$^\circ$C is 2461.5 days (6.7 years).

\begin{table}
\centering
\caption{Overview of dark current levels and classification of SiPM tiles based on goodness-of-fit and temperature correlation}
\label{tab:cat-current}
\begin{tabular}{|l|c|c|c|c|c|}
\hline
Category \textbackslash\, Dark Current ($\mu$A) & [0, 30) & [30, 60) & [60, 100) & $\geq$100 & Total\\
\hline
\multicolumn{6}{|c|}{Fit Success ($\chi^2 \leq$ 6.5 )} \\
\hline
Normal & 0 & 3597 & 1 & 0 & 3598 \\
\hline
\multicolumn{6}{|c|}{Fit Failure ($\chi^2 >$ 6.5 )} \\
\hline
w/ Temperature Correlation ($R_S$ $\geq$ 0.8) & 1 & 347 & 1 & 11 & 360\\
w/o Temperature Correlation ($R_S$ $<$ 0.8) & 0 & 30 & 27 & 2  & 59\\
\hline
Total & 1 & 3974 & 29 & 13 & 4017\\
\hline
\end{tabular}

\end{table}

\section{Summary}
\label{sec:summary}
This study presents the findings from both the SiPM burn-in test and characterizations within a cryogenic environment. Although the burn-in data (current, temperature, time) are limited in scope, we successfully derived meaningful insights using a specialized dark current model, incorporating several critical parameters. The dark current model effectively fits the burn-in data and yields acceptable chi-square ($\chi^2$) values for most tiles. We also employed a robust approach to examine the current-temperature correlation with Spearman coefficients to account for the fit failures. A temperature correction is also developed to account for potential thermal inertia. Additionally, we discovered that the breakdown voltage, determined through the burn-in dark current model, aligns closely with traditional methods that extrapolate based on temperature.

Furthermore, we assess the dark current at room temperature of more than 4,000 SiPM tiles, which together provide a photon-sensitive surface area of nearly 10 m$^2$. 
Our analysis finds that the majority of the SiPM tiles exhibit dark current levels ranging between 30 to 60 $\mu$A. Nonetheless, we identify anomalies among the 4,000 SiPM tiles. We find 13 tiles with dark current values exceeding 100 $\mu$A, higher than the norm, whereas 59 tiles demonstrated poor correlation between dark current and temperature. These anomalous SiPM tiles are slated for further inspection at the cryogenic testing facility to understand how their peculiar behavior during the burn-in test affects key performance parameters. Despite the presence of anomalies, all SiPMs successfully passed the two-week burn-in test at room temperature, which is equivalent to 6.7 years at -50°C, satisfying the expected operation time of 6 years in Taishan.

\acknowledgments
We gratefully acknowledge the support from the RSF-NSFC Cooperation Funding under Grant No. 12061131008 and the National Key Research and Development Program of China under Grant No. 2022YFA1602002. 


\clearpage
\bibliographystyle{JHEP}
\bibliography{biblio.bib}

\end{document}